\documentstyle[12pt,epsf]{article}
\begin{document}

\def\beq{\begin{equation}}
\def\eeq{\end{equation}}  
\def\bea{\begin{eqnarray}}
\def\eea{\end{eqnarray}}  
\def\bq{\begin{quote}}    
\def\eq{\end{quote}}      
\def\ra{\rightarrow}      
\def\lra{\leftrightarrow} 
\def\ups{\upsilon}
\def\nn{\nonumber}
\def\r.{\right.}
\def\l.{\left.}
\def\o.{\overline}
\newcommand{\sheptitle}
{String Unification, Spaghetti Diagrams and Infra-Red Fixed Points}
\newcommand{\shepauthor}
{B. C. Allanach$^1$ and
S. F.
King$^2$ \\ 
\vspace{\baselineskip}
{\small 1. Rutherford Appleton Laboratory, Chilton, Didcot, OX11 0QX, U.K.} \\
{\small 2. Department of Physics and Astronomy, University of
Southampton, Southampton, SO17 1BJ, U.K.}}

\newcommand{\shepabstract}
{The minimal supersymmetric standard model (MSSM),
is perhaps the leading candidate for new physics beyond the
standard model, but it encounters difficulties with string 
gauge unification and in addition does not shed any light on the
question of fermion masses.
We consider a scenario in which the MSSM
is valid up to an energy scale of 
$\sim 10^{16}$ GeV, but that above this scale the theory is 
supplemented by extra 
vector-like representations of the gauge group,
plus a gauged $U(1)_X$ family symmetry.
In our approach the extra heavy matter
above the scale $\sim 10^{16}$ GeV is used in two different
ways: (1) to allow (two-loop) gauge 
coupling unification at the string scale; (2) to mix with
quarks, leptons and Higgs fields via spaghetti diagrams
and so lead to phenomenologically
acceptable Yukawa textures. 
We determine the most economical models in which the
extra matter can satisfy both constraints simultaneously.
We then give a general discussion of the
infra-red fixed points of such models, pointing out the conditions for
infra-red stability, then discuss two semi-realistic examples:
a Higgs mixing model, and a quark mixing model.}

\begin{titlepage}
\begin{flushright}
RAL 97-007 \\ SHEP 97-01 \\ {\tt hep-ph/9703293}
\end{flushright}
\vspace{.4in}
\begin{center}
{\large{\bf \sheptitle}}
\bigskip \\ \shepauthor \\ \mbox{} \\
\vspace{.5in}
{\bf Abstract} \bigskip \end{center} \setcounter{page}{0}
\shepabstract
\end{titlepage}

\section{Introduction}

The apparent 
unification of the gauge couplings 
at a scale $M_{GUT}\sim 10^{16}$ GeV \cite{GUTun}
is an encouraging feature of the MSSM\@,
whose direct experimental verification so far remains out of reach.
However the MSSM does not address the question of fermion masses,
and in any case it is unlikely that the MSSM can survive unchanged above
$M_{GUT}\sim 10^{16}$ GeV since the gauge couplings which converge on this
scale begin to diverge above it, and are quite unequal at the string scale
$M_X\sim 5\times 10^{17}$ GeV, even taking into account higher
Kac-Moody levels and string threshold effects \cite{Dienes}.
The traditional approach is to embed the MSSM in some
supersymmetric grand unified
theory (SUSY GUT) but such an 
approach presents many theoretical and phenomenological
challenges \cite{HR,su5xsu5,level2},
and we shall not consider it further here.

String models predict that the three gauge
couplings of the MSSM
are directly related to each other at the string scale $M_X$
\cite{gaugeun}.
Weakly coupled string theories give the relation \cite{kap}
\begin{equation}
M_X = 5.27 \times 10^{17} g_X \mbox{~GeV},
\label{stringscale}
\end{equation}
where $g_X$ is the unified gauge coupling at the string scale $M_X$.
If the MSSM (and nothing else) persists
right up to the string scale $M_X$ such theories
are in conflict with low energy measurements of the gauge couplings. From this
data, one can show that
gauge couplings cross at $\sim 10^{16}$ GeV,
and significantly diverge at the string scale
$\sim 5\times 10^{17}$ GeV. However the situation is in fact not 
so clear cut since the $U(1)_Y$ hypercharge gauge coupling
has an undetermined normalisation factor $k_1\geq 1$ 
(where for example $k_1=5/3$ is the usual GUT normalisation)
which may be set to be a phenomenologically desired value
\cite{ibanez} by the choice of a particular string model. However the
simplest string theories
(e.g.\ heterotic string with any standard compactification)
predict equal gauge couplings for the other two observable sector
gauge groups
$g_2=g_3$ at the string scale $M_X$, which would require
a rather large correction in order to
account for $\alpha_s(m_Z)$~\cite{kap,thresholds}. In fact,
a recent analysis \cite{Dienes,Dinesetal}
concludes that string threshold effects are insufficient by themselves
to resolve the experimental discrepancy. The analysis also concludes
that light SUSY thresholds and two-loop corrections
cannot resolve the problem, even when acting together.
In order to allow the gauge couplings to unify at the string scale
it has been suggested \cite{ellisetc} that additional heavy exotic matter
in vector-like representations should
be added to the MSSM at some intermediate scale or scales $M_I<M_X$,
leading to the so called MSSM+X models.
A detailed unification analysis of 
such models was performed by Martin and Ramond
(MR) \cite{MR} for example.
In a previous paper \cite{MSSM+X} we
performed a general unification analysis of
MSSM+X models, focusing on the 
infra-red fixed point properties of the top quark mass prediction,
using similar techniques to those proposed for the
MSSM and GUTs \cite{PR,H,RLfixed}. 
The main result was that the top quark mass tends to be
heavier than in the MSSM, and closer to its quasi-fixed point
in these models. 

This leads us to the question of fermion masses in the MSSM\@.
The fermion mass problem in the MSSM arises from the
presence of many unknown dimensionless Yukawa couplings.
A possible solution to this long-standing problem is that the
MSSM could be embedded in some more predictive high energy
theory with fewer Yukawa couplings. Alternatively one may attempt to
relate the Yukawa couplings to the gauge coupling,
as in the Pendleton-Ross infra-red stable
fixed point (IRSFP) for the top quark Yukawa coupling \cite{PR},
or the quasi-fixed point of Hill \cite{H}\footnote{The relation between
the fixed point and
quasi-fixed point has been fully elucidated by Lanzagorta and Ross
\cite{RLfixed}.
The conditions for the stability
of fixed points for any number of dimensionless Yukawa couplings in general 
supersymmetric models has also recently been examined
\cite{gem}.}.
To obtain IRSFPs it is necessary that the unknown   
dimensionless Yukawa couplings are of the same order
as the gauge coupling at high energy. Thus at first sight this
approach would seem to be inapplicable to the small Yukawa couplings
of the standard model.
However string theory yields large Yukawa couplings of order the
gauge coupling, and this
could suggest that the small dimensionless
Yukawa couplings be reinterpreted as large Yukawa couplings
multiplied by a ratio of mass scales. In such a scenario   
fixed points may once again be applicable, as pointed out recently by
Ross \cite{GG}. In this case the high energy scale would be the 
string scale and the infra-red region would be near the GUT scale.   
Thus IRSFPs may provide an unexpected resolution of the
fermion mass problem. 

In the present paper we consider a bottom-up string-inspired 
approach in which 
the MSSM is valid up to an energy scale of 
$M_I \sim 10^{16}$ GeV. Above this scale the theory is 
supplemented by extra vector-like representations of the gauge group,
plus a gauged $U(1)_X$ family symmetry, which turns out to be
pseudo-anomalous as discussed in the next section.
Although we do not derive our results from a string model,
both the presence of extra $U(1)$ gauge groups
(one of which is pseudo-anomalous and 
may be identified with the gauged family symmetry $U(1)_X$),
and extra vector-like matter, are typical of recent string
compactifications\footnote{For more discussion about this point see
ref.\cite{george}.}.
The basic idea of our present approach is that the extra heavy matter
above the scale $\sim 10^{16}$ GeV may be used in two different
ways: (1) to allow (two-loop) gauge 
coupling unification at the string scale; (2) to mix with
quarks, leptons and Higgs fields via spaghetti diagrams
and so lead to phenomenologically
acceptable Yukawa textures. We emphasise that
in our approach the operators required for Yukawa textures
are generated from the dynamics of the effective field
theory beneath the string scale rather than from the string theory itself.
The overall philosophy of the approach thus far is to explain the data on
fermion masses and mixing angles {\em qualitatively}. 
Unfortunately, once the extra
fields have been added there are many extra free parameters than data
points. In order to increase the predictivity of the class of models
we give a detailed discussion of the IRSFP properties of the models.
As indicated above, the IRSFP approach has
the potential to predict all of the masses and mixings in terms of about two
free parameters, although as we shall see there are difficulties
in practice with implementing such a scheme.

In this paper we essentially build on the work of ref.\cite{GG},
where a specific model with a gauge $U(1)_X$ family symmetry
was introduced, and the extra vector-like matter that was introduced
consisted of extra Higgs doublets which mixed with the ordinary
Higgs doublets.
Our present work is more general and differs in several respects.
For example the question of gauge coupling unification
was addressed in ref.\cite{GG}
by adding complete $SU(5)$ $5 \oplus \bar{5}$ representations 
(which contain the additional Higgs states 
required for the mixings)
to the MSSM theory with masses just below the unification scale.
These have no relative effect on the running of the three gauge couplings
to one loop order, however at two-loop order
it was claimed that the unification
scale is raised. In our two loop
analysis we explicitly find that such 
a mechanism does not increase the unification scale sufficiently,
so we abandon this possibility here. Instead we perform a 
general two-loop analysis of string gauge unification,
including heavy matter at a scale $M_I$ which consists of
MSSM vector representations which do not form complete
$SU(5)$ representations, i.e.\ we allow split $SU(5)$ representations.
In this respect our approach resembles that in refs.\cite{ellisetc,MR,MSSM+X},
however we emphasise that in our present approach we impose the 
additional constraint that the intermediate mass scale $M_I$
must be very close to the string scale $M_X$ because, as
we shall see in the next section,
the $U(1)_X$ gauge symmetry is pseudo-anomalous
and must be broken not far below the scale $M_X$. 
In practice this implies that $M_I \sim 10^{16}$ GeV,
as stated earlier.
Another difference between
our work here and ref.\cite{GG} is that we
allow more general mixing possibilities
along the Higgs, quark and lepton lines of the spaghetti diagrams,
rather than just the Higgs mixing considered in ref.\cite{GG}. 
Finally we discuss the IRSFPs in this more general framework,
rather than just the particular Higgs mixing model in ref.\cite{GG}.

We emphasise that 
in our approach the extra states required for unification are 
also used for spaghetti mixing. This economical double use of the
extra heavy matter is the main new idea of the present paper,
and one of our main results is to tabulate the minimal 
amount of extra vector-like matter which allows {\em both}
\/string gauge unification (subject to the high $M_I$ constraint) 
and phenomenologically acceptable Yukawa textures
(see Tables~\ref{tab:mainresults} and~\ref{tab:res2}.) 

The layout of the remainder of this paper is as follows.
In section~2 we show how Yukawa textures may be generated from
a broken $U(1)_X$ gauged family symmetry where the desired operators
are generated from the effective field theory beneath the string scale
via spaghetti diagrams. In section~3 we discuss a 
two-loop analysis of string gauge unification,
where the extra states are consistent with the 
requirements of the previous section.
In section~4 we give a general discussion of the infra-red fixed point
nature of such models, then discuss
two semi-realistic examples.
Section~5 concludes the paper. 
Appendix~1 summarises the running of the gauge couplings to two loops in
models with matter additional to the MSSM\@, appendix~2 lists the
wave-function renormalisations for the general model discussed in section~4,
appendix~3 details the infra-red fixed point of the Higgs mixing model,
and appendix~4 details the infra-red fixed point of a quark-line mixing model.

\section{Yukawa Textures from
$U(1)_X$ Family Symmetry, Vector Representations and Spaghetti Diagrams}

The pattern of quark and lepton masses and quark mixing angles
has for a long time been a subject of fascination for particle physicists.
In terms of the standard model, this pattern arises from
three by three complex Yukawa matrices (54 real parameters)
which result in nine real eigenvalues plus four real mixing parameters
(13 real quantities) which can be measured experimentally.
In recent years the quark and lepton
masses and mixing angles have been measured with increasing
precision, and this trend is likely to continue in the future as
lattice QCD calculations provide increasingly accurate estimates
and B-factories come on-line.
Theoretical progress is less certain, although there has been a steady
input of theoretical ideas over the years and in recent times there is
an explosion of activity in the area of supersymmetric unified models.
This approach presumes that at very high energies close to the unification
scale,
the Yukawa matrices exhibit a degree of simplicity, with simple relations
at high energy corrected by the effects of renormalisation group (RG)
running down to low energy. For example, the idea of bottom-tau,
top-bottom-tau and even top-bottom-tau-tau-neutrino
Yukawa unification have
received a good deal of attention recently (see for example 
ref.\cite{quadruple}.)

These successes with the third family relations are not
immediately generalisable to the lighter families. For the remainder
of the Yukawa matrices, additional ideas are required in order
to understand the rest of the spectrum. One such idea is that of
texture zeroes: the idea that the Yukawa matrices at the unification
scale are rather sparse; for example the Fritzsch ansatz \cite{Fritzsch}.
Although the Fritzsch texture does not work for
supersymmetric unified models, there are other textures which do, for
example the Georgi-Jarlskog (GJ) texture \cite{GJ}
for the down-type quark and lepton matrices:
\begin{equation}
\lambda^E = \left(\begin{array}{ccc}
0 & \lambda_{12} & 0 \\
\lambda_{21} & - 3 \lambda_{22} & 0 \\
0 & 0 & \lambda_{33} \\ \end{array}\right)
,\ \
\lambda^D = \left(\begin{array}{ccc}
0 & \lambda_{12} & 0 \\
\lambda_{21} & \lambda_{22} & 0 \\
0 & 0 & \lambda_{33} \\ \end{array}\right).
\label{GJ}
\end{equation}
After diagonalisation this leads to
$\lambda_{\tau} = \lambda_b$,
$\lambda_{\mu} = 3\lambda_s$,
$\lambda_e = \lambda_d/3$
at the scale $M_{GUT}$ which result in (approximately)
successful predictions at low energy. Actually the factor of 3 in the 22
element above arises from group theory: it is a Clebsch factor coming from
the choice of Higgs fields coupling to this element.

It is observed that if we choose the upper two by two block
of the GJ texture
to be symmetric, $\lambda_{12}=\lambda_{21}$, and if we can disregard
contributions from the up-type quark matrix, then we also have the successful
mixing angle prediction
\begin{equation}
V_{us}=\sqrt{\lambda_d/\lambda_s}.
\end{equation}
The data therefore supports the idea of symmetric matrices, and a texture
zero in the 11 position. Motivated by the desire for maximal predictivity,
Ramond, Roberts and Ross (RRR) \cite{RRR} have made a survey of possible  
symmetric textures which are both consistent with data and involve
the maximum number of texture zeroes. Assuming GJ relations for the
leptons, RRR tabulated five possible
solutions for the up-type and down-type Yukawa matrices.

Having identified successful textures\footnote{Over the recent years,
there has been an extensive study of fermion mass
matrices with zero textures \cite{oth1}.
}, the obvious questions are:
what is the origin of the texture zeroes? and: what is the origin
of the hierarchies (powers of the expansion parameter $\epsilon$)?
A natural answer to both these questions was provided early on   
by Froggatt and Nielsen (FN) \cite{FN}. 
A specific realisation of the FN idea was
provided by Ibanez and Ross (IR) \cite{IR}, based on the MSSM extended
by a gauged family $U(1)_X$ symmetry with $\theta$ and $\bar{\theta}$
singlet fields with opposite $X$ charges, plus new heavy Higgs fields
in vector representations\footnote{The generalisation to include
neutrino masses
is straightforward \cite{DLLRS}.
}. Anomaly cancellation
occurs via a Green-Schwarz-Witten
(GSW) mechanism, and the $U(1)_X$ symmetry is broken not far below the 
string scale \cite{IR}. 
The idea is that the $U(1)_X$ family
symmetry only allows the third family to receive a renormalisable
Yukawa coupling
but when the family symmetry is broken at a scale not far below the
string scale other families receive suppressed effective Yukawa
couplings.
The suppression factors are proportional to powers of
the vacuum expectation values (VEVs) of $\theta$ fields which are MSSM
singlets but carry
$U(1)_X$ charges and are
responsible for breaking the family
symmetry. The non-renormalisable 
operators that give the small effective Yukawa
couplings are scaled by heavier mass scales $M_I$ 
identified as the
masses of new heavy vector representations which also carry
$U(1)_X$ charges. IR envisaged a series of
heavy Higgs doublets of mass $M_I$
with differing $U(1)_X$ charges 
which couple to the lighter families via sizable Yukawa
couplings that respect the family symmetry.
The heavy Higgs doublets also couple to the MSSM Higgs doublets
via $\theta$ fields and this results in suppressed effective   
Yukawa couplings when the family symmetry is broken.

By making certain symmetric charge assignments (see later),
IR showed that the RRR texture solution 2
could be approximately reproduced. To be specific, for a certain choice
of $U(1)_X$ charge assignments, IR generated Yukawa matrices of the
form:
\begin{equation}
\lambda^U = \left(\begin{array}{ccc}
\epsilon^8 & \epsilon^3 & \epsilon^4 \\
\epsilon^3 & \epsilon^2 & \epsilon \\  
\epsilon^4 & \epsilon & 1 \\ \end{array}\right)
,\ \
\lambda^D = \left(\begin{array}{ccc}
\bar{\epsilon}^8 & \bar{\epsilon}^3 & \bar{\epsilon}^4 \\
\bar{\epsilon}^3 & \bar{\epsilon}^2 & \bar{\epsilon} \\  
\bar{\epsilon}^4 & \bar{\epsilon} & 1 \\ \end{array}\right)
,\ \
\lambda^E = \left(\begin{array}{ccc}
\bar{\epsilon}^5 & \bar{\epsilon}^3 & 0 \\
\bar{\epsilon}^3 & \bar{\epsilon} & 0 \\  
0  & 0  & 1 \\ \end{array}\right)
\label{IR}
\end{equation}
These are symmetric in the expansion parameters $\epsilon$ and
$\bar{\epsilon}$, which are regarded as independent parameters.
This provides a neat and predictive framework, however there are
some open issues.
Although the order of the entries is fixed by the expansion parameters,
there are additional
parameters of order unity multiplying each entry, making precise predictions
difficult. A way to address the problem of the unknown
coefficients has been proposed in
\cite{GG} where it has been shown that the various coefficients
may arise as a result of the infra-red fixed-point structure
of the theory beyond the Standard Model. The idea behind this
approach is that since there are no small Yukawa couplings
one may hope to determine all the Yukawa couplings
by the use of infra-red fixed points along similar lines
to the top quark Yukawa coupling determination. This attractive idea thus
allows the determination of otherwise unknown parameters in the model, simply
by the dynamics of the renormalisation group (RG) flow of the model,
as discussed in Section 4.

Having given an overview of the approach, we now discuss in a little
more detail how the $U(1)_X$ gauge symmetry is introduced.
The way this is achieved is well
documented \cite{IR} and here we only sketch the main results.
The quark and lepton multiplets are assigned 
family dependent (FD) charges 
as shown in Table~\ref{Table1}.

\begin{table}[h]
\centering
\begin{tabular}{|c |ccccccc|}\hline
   &$ Q_i$ & $U^c_i$ &$ D^c_i$ &$ L_i$ & $E^c_i$ & $H_1$ & $ H_2$   \\
\hline
  $U(1)_{FD}$ & $\alpha _i$ & $\alpha _i$  & $\alpha_i$
& $\alpha_i$ & $\alpha_i $ & $-2\alpha _3$ &  $-2\alpha _3$
\\
\hline
\end{tabular}
\caption{{\small $U(1)_{FD}$ charges assuming symmetric textures.}
\label{Table1}}
\end{table}

The need to preserve $SU(2)_L$
invariance requires left-handed up and down quarks (leptons)
to have the same charge. This, plus the additional
requirement of symmetric
matrices, indicates that all quarks (leptons) of the same i-th
generation transform with the same charge $\alpha _i$.
It is further assumed that
quarks and leptons of the same family have the same charge
(this additional assumption was made by Ross \cite{GG}).
The full anomaly free Abelian group involves an additional family
independent component, $U(1)_{FI}$, and with this freedom
$U(1)_{FD}$ is made traceless without any loss of
generality\footnote{Since we assume that the 33 operator is
renormalisable, the relaxation of the
tracelessness condition does not change the
charge matrix since any additional FI charges can always be absorbed
into the Higgs charges $H_{1,2}$.}.
Thus we set $\alpha_1=-(\alpha_2+\alpha_3)$. 

Making the above assumptions all charge and mass
matrices have the same structure under the
$U(1)_{FD}$ symmetry.
The FD charge matrix involving two quark or lepton fields and a Higgs
is of the form
\begin{eqnarray}
\left(
\begin{array}{ccc}
-2 \alpha_2 - 4 \alpha_3 &
-3 \alpha_3 & -\alpha_2 - 2\alpha_3 \\
-3 \alpha_3 & 2(\alpha_2-\alpha_3)
& \alpha_2 - \alpha_3 \\
-\alpha_2 -2 \alpha_3 & \alpha_2 - \alpha_3 & 0
\end{array}
\right)
\label{chargematrix}
\end{eqnarray}
Acceptable textures are obtained for
\beq
\frac{\alpha_3}{\alpha_2 - \alpha_3}=1
\eeq
or
\beq
{\alpha_2 }=2{\alpha_3}
\eeq

However these are only the family dependent charges.
The total $U(1)_X$ charges are given by
\beq
U(1)_X=U(1)_{FD}+U(1)_{FI}
\eeq
where
\beq
U(1)_{FI}\equiv U(1)_{T}+U(1)_{F}
\eeq
where the corresponding FI charges are denoted by $t,f$ 
and the resulting $U(1)_X$ charges are given in Table~\ref{Table2}.

\begin{table}[h]
\centering
\begin{tabular}{|c |ccccccc|}\hline
   &$ Q_i$ & $U^c_i$ &$ D^c_i$ &$ L_i$ & $E^c_i$ & $H_1$ & $ H_2$   \\
\hline
  $U(1)_{X}$ & $\alpha _i+t$ & $\alpha _i+t$  & $\alpha_i+f$
& $\alpha_i+f$ & $\alpha_i+t$ & $-2\alpha _3-(f+t)$ &  $-2\alpha _3-2t$
\\
\hline
\end{tabular}
\caption{{\small $U(1)_{X}$ charges assuming symmetric textures.}
\label{Table2}}
\end{table}

At this point we should be clear about the question
of anomalies for the various $U(1)$'s which were introduced into this model
\cite{IR}. We began by introducing a $U(1)_{FD}$ symmetry,
which was arranged to be traceless without loss of generality.
Although the tracelessness is sufficient to guarantee that the quarks
and leptons contribute a zero anomaly, the phenomenologically required
Yukawa texture implies that the Higgs doublets must carry a
$U(1)_{FD}$ charge (the above conditions cannot be satisfied by $\alpha_3=0$)
and consequently the Higgs contribute a non-zero $SU(2)^2U(1)_{FD}$ anomaly.
In order to improve matters, we returned to the $U(1)_{FI}$ symmetry
which is a family independent symmetry, identified as the
sum of $U(1)_F$ and $U(1)_T$. Although there is no
explicit $SU(5)$ unification, the components of the 
``5'' rep are assigned a charge $f$ under $U(1)_F$, 
while the components of the
``10'' are assigned a charge $t$ under $U(1)_T$, 
and the Higgs doublets are assigned the
charges which guarantee that the Yukawa charge matrix remains unchanged
under the total $U(1)_X$ symmetry defined above
as the sum of the $U(1)_{FD}$ and $U(1)_{FI}$ symmetries.
Although, following ref.\cite{IR}, we have introduced several $U(1)$'s,
it should be remembered that at the end of the day only one of them
(the total $U(1)_X$) is regarded as the physical gauged family
symmetry, and the others are for construction purposes only.
The essential point to note is that the choice of FI charges in 
Table~\ref{Table2} is the most general 
choice consistent with anomaly cancellation 
amongst the quarks and leptons via the Green-Schwarz mechanism
\cite{GS} which is based on assigning $SU(5)$ multiplets
equal charges. The physics of the GS mechanism in 4D is that
one cancels the anomaly of the $U(1)_X$ by an appropriate
shift of the axion present in the dilaton multiplet
of 4D strings. For the GS mechanism to be possible, the coefficients
of the mixed anomalies of the $U(1)_X$ with $SU(3)$,
$SU(2)$ and $U(1)_Y$ have to be in the ratio of the Kac-Moody levels
$k_3:k_2:k_1$, and hence the GS mechanism works if the
anomalies are in the ratio $1:1:5/3$, and this is in fact guaranteed
by the choice of quark and lepton charges in Table~\ref{Table2}.
This of course still leaves the question of the contribution
to the anomalies from the Higgs doublets.
In order to allow anomaly cancellation
amongst the Higgs doublets, and permit a renormalisable $\mu$ mass
term between the Higgs doublets we must arrange that the Higgs carry
zero $X$ charge, which in turn implies that the third family has zero 
$X$ charge, and this is accomplished by choosing:
\beq
f=t=-\alpha_3, \label{chargecond}
\eeq
Putting $\alpha_3=1$, where the $U(1)_X$ symmetry is broken by the VEVs of
MSSM singlet fields $\theta$ and $\bar{\theta}$ with 
$U(1)_X$ charges -1 and +1 respectively
implies the following simple $X$ charges for the three families:
\begin{equation}
\begin{array}{cc}
1^{st} \ \mbox{family:} &  X=-4 \\
2^{nd} \ \mbox{family:} &  X=+1 \\
3^{rd} \ \mbox{family:} &  X=0  \\
\end{array}.
\end{equation}
Since the Higgs charges are zero the charge matrix is simply the
matrix given by the sum of the quark (or lepton) charges
from each family:
\begin{eqnarray}
\left(
\begin{array}{ccc}
-8 & -3  & -4\\
-3 & +2  & +1\\
-4 & +1  & 0
\end{array}
\right)
\label{chargematrix2}
\end{eqnarray}
In any mass term, the sum of the charge of the fields in that term must be
zero to preserve U(1)$_X$. 
Thus from Eq.\ref{chargematrix2}, the operators that could possibly generate
the $U$ quark mass matrix are:
\bea
Q_3U^c_3H_2,\ & Q_3U^c_2H_2(\bar{\theta}),\ & Q_2U^c_3H_2(\bar{\theta}),
\nn \\
Q_2U^c_2H_2(\bar{\theta})^2,\ & Q_3U^c_1H_2({\theta})^4,
\ & Q_1U^c_3H_2({\theta})^4,\nn \\
Q_1U^c_1H_2({\theta})^8,\ & Q_2U^c_1H_2({\theta})^3,
\ & Q_1U^c_2H_2({\theta})^3.\label{nonrenops}
\eea
Because each $\theta$ or $\bar{\theta}$ field corresponds
to a suppression factor of
$\epsilon$ (where $\epsilon=<\theta >/M_I=<\bar{\theta}>/M_I$ )
the corresponding texture is of order
\begin{equation}
\left(
\begin{array}{ccc}
\epsilon^8 & \epsilon^3  & \epsilon^4\\
\epsilon^3 & \epsilon^2  & \epsilon\\
\epsilon^4 & \epsilon  & 1
\end{array}
\right).
\label{epsilonmatrix}
\end{equation}

We now turn to the question of the origin of the non-renormalisable
operators in Eq.\ref{nonrenops}.
A natural answer to this question was provided   
by Froggatt and Nielsen (FN) \cite{FN}. The basic idea involves
some new heavy matter of mass\footnote{We will identify the new heavy
matter
required to generate the 
non-renormalisable operators with the new heavy matter required
to ensure string gauge unification so we take the mass scale to be $M_I$.}
$M_I$
which are in vector representations of the MSSM gauge group
and which carry charges under the family group $U(1)_X$.
The vector-like matter couples to ordinary
matter (quarks, leptons, Higgs) via the 
MSSM singlet fields $\theta$ and $\bar{\theta}$
leading to ``spaghetti-like'' tree-level diagrams. 
The spaghetti diagrams yield the effective non-renormalisable      
operators in Eq.\ref{nonrenops}.

\begin{figure}
\begin{center}
\leavevmode   
\hbox{\epsfxsize=3in
\epsfysize=2in
\epsffile{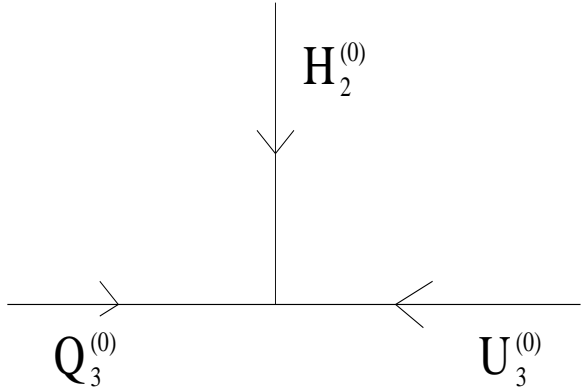}}
\end{center}
\caption{Renormalisable 33 operator.}
\label{Fig1}
\end{figure}

An explicit realisation of this mechanism was discussed by
IR\cite{IR} and subsequently by Ross \cite{GG}, based on 
the heavy vector-like matter corresponding to additional
Higgs doublets which could mix with the MSSM doublets
$H_1,\ H_2$  via the spaghetti diagrams.
Thus the following Higgs were introduced \cite{GG},
\begin{equation}
H_{1,2}^{(-1)}, \bar{H}_{1,2}^{(1)},
H_{1,2}^{(-2)}, \bar{H}_{1,2}^{(2)},
H_{1,2}^{(3)}, \bar{H}_{1,2}^{(-3)},
H_{1,2}^{(4)}, \bar{H}_{1,2}^{(-4)},
H_{1,2}^{(8)}, \bar{H}_{1,2}^{(-8)}, \label{insanity}
\end{equation}
where the $U(1)_X$ charges are given in parentheses,
and $H_{1,2}^{(x)}$ have hypercharges $Y/2=-1/2,1/2$,
and $\bar{H}_{1,2}^{(-x)}$ have hypercharges $Y/2=1/2,-1/2$,
respectively. The idea is that the Higgs $H_{1,2}^{(x)}$
have direct couplings to the lighter families and
mix with the MSSM Higgs $H_{1,2}$ via singlet
$\theta$ fields. Thus the {\em renormalisable} \/Higgs
terms (where we have neglected the Yukawa couplings) are :
\begin{equation}
\left( \begin{array} {ccc}
Q_1 & Q_2 & Q_3 \\ \end{array} \right)
\left(
\begin{array}{ccc}
H_{1,2}^{(8)} & H_{1,2}^{(3)} & H_{1,2}^{(4)}\\
H_{1,2}^{(3)} & H_{1,2}^{(-2)}  & H_{1,2}^{(-1)}\\
H_{1,2}^{(4)} & H_{1,2}^{(-1)} & H_{1,2}
\end{array}
\right)
\left(
\begin{array}{c} U_1^c \\ U_2^c \\ U_3^c \\ \end{array} \right).
\label{Higgsmatrix}
\end{equation}
For example, the renormalisable 33 operator is depicted in Fig.\ref{Fig1}.

\begin{figure}
\begin{center}
\leavevmode   
\hbox{\epsfxsize=3in
\epsfysize=3in
\epsffile{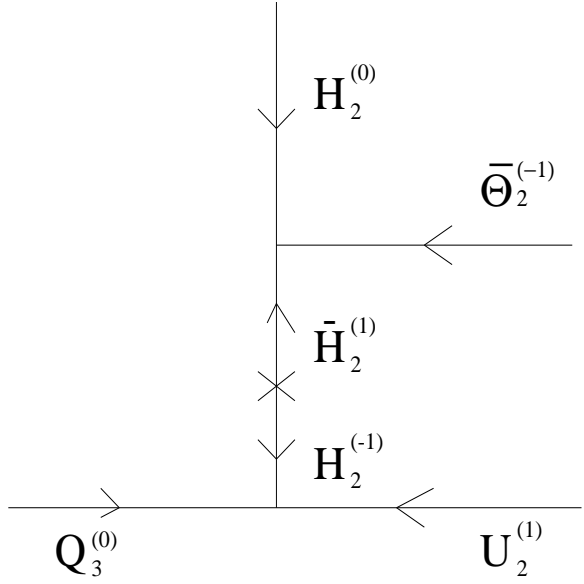}}
\end{center}
\caption{Spaghetti diagram for 32 mixing}
\label{Fig2}
\end{figure}

\begin{figure}
\begin{center}
\leavevmode   
\hbox{\epsfxsize=3in
\epsfysize=5in
\epsffile{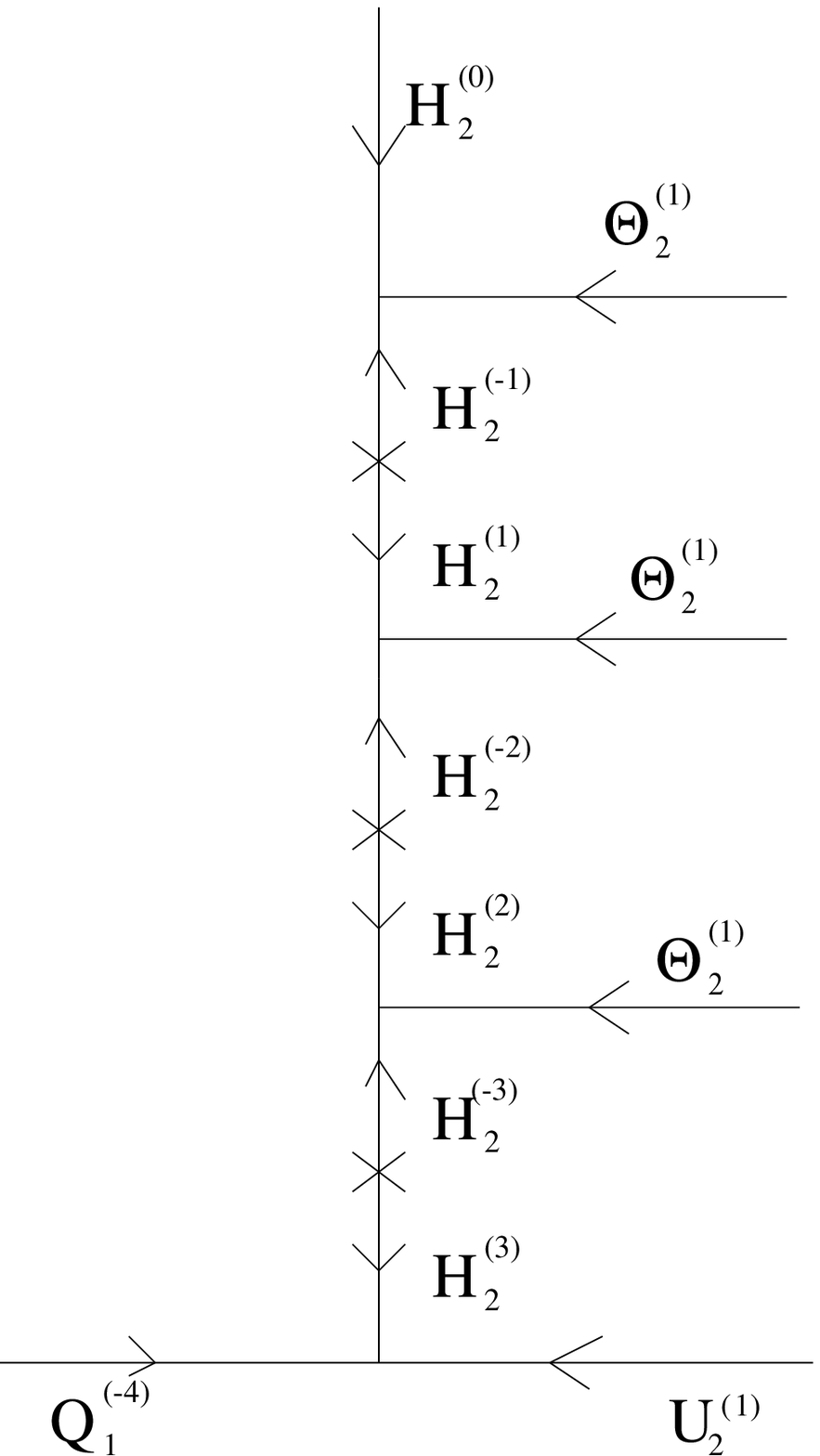}}
\end{center}
\caption{Spaghetti diagram for Cabibbo (12) mixing}
\label{Fig3}
\end{figure}

Such direct Higgs couplings, allowed by the $U(1)_X$ symmetry,
combined with Higgs mixing via singlet field insertions, 
lead to the effective non-renormalisable operators in Eq.\ref{chargematrix2}.
For example the spaghetti diagram responsible for the 32 quark mixing term
is illustrated in Fig.\ref{Fig2}.
It is clear that such diagrams generate all the operators in 
Eq.\ref{nonrenops}. By drawing such diagrams
it becomes clear that 
additional heavy vector Higgs are required beyond those
in the direct coupling matrix in Eq.\ref{Higgsmatrix}.
It is easy to see that all Higgs charges in integer steps
between 8 and -2 are required if all elements of the mixing 
matrix are to be non-zero. For example Cabibbo mixing requires
the additional Higgs with charges 2 and 1 (plus their conjugates)
so that the Higgs of charge 3 can step down to the Higgs of zero charge
via three $\theta$ field insertions as shown in Fig.\ref{Fig3}.

\begin{figure}
\begin{center}
\leavevmode   
\hbox{\epsfxsize=5in
\epsfysize=3in
\epsffile{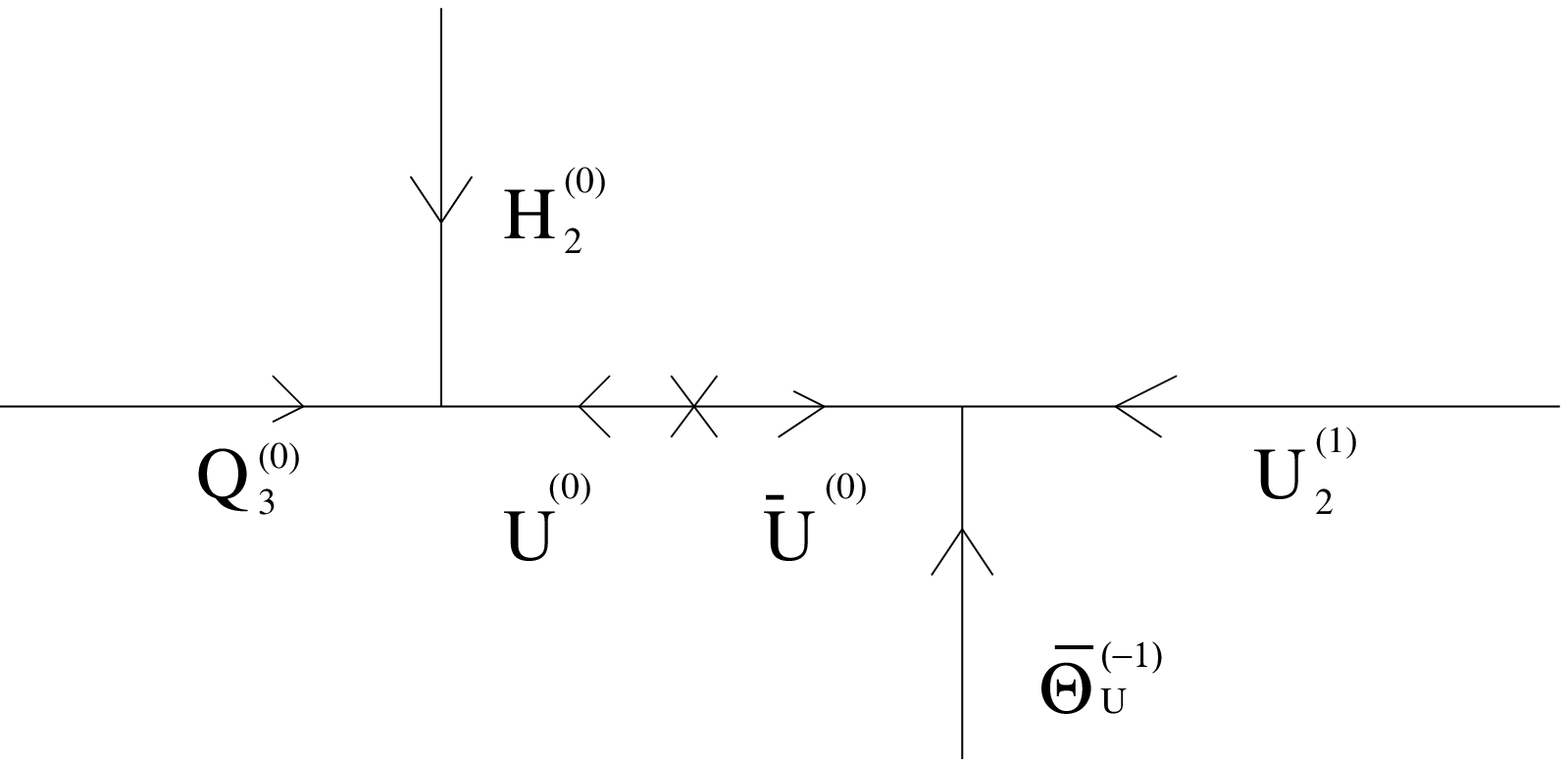}}
\end{center}
\caption{Spaghetti diagram for 32 mixing along the U line}
\label{Fig4}
\end{figure}

\begin{figure}
\begin{center}
\leavevmode   
\hbox{\epsfxsize=5in
\epsfysize=3in
\epsffile{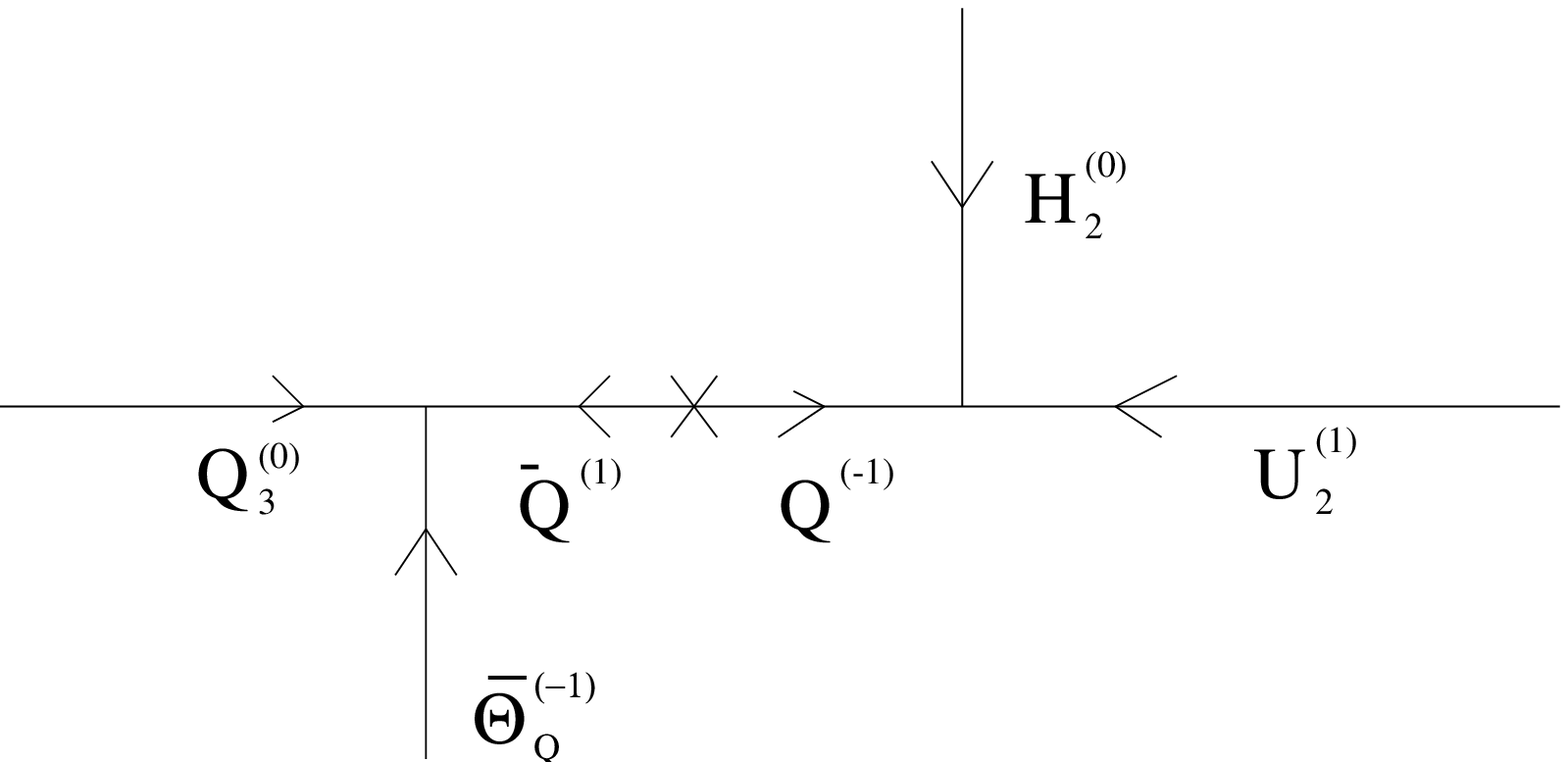}}
\end{center}
\caption{Spaghetti diagram for 32 mixing along the Q line}
\label{Fig5}
\end{figure}

The full list of Higgs required for this scenario is
larger than assumed in Eq.\ref{insanity} and is displayed below:
\bea
  H_{1,2}^{(8)}, \bar{H}_{1,2}^{(-8)}, 
 H_{1,2}^{(7)},  \bar{H}_{1,2}^{(-7)},
  H_{1,2}^{(6)},  \bar{H}_{1,2}^{(-6)},
 H_{1,2}^{(5)}, \bar{H}_{1,2}^{(-5)},
  H_{1,2}^{(4)},  \bar{H}_{1,2}^{(-4)},\nn \\
 H_{1,2}^{(3)},  \bar{H}_{1,2}^{(-3)},
  H_{1,2}^{(2)},  \bar{H}_{1,2}^{(-2)},
 H_{1,2}^{(1)}, \bar{H}_{1,2}^{(-1)},
  H_{1,2}^{(-1)},  \bar{H}_{1,2}^{(1)},
 H_{1,2}^{(-2)}, \bar{H}_{1,2}^{(2)} 
\label{fullHiggs}
\eea
Since complete Higgs mixing requires 20 Higgs vector representations,
rather than 10 as assumed on the basis of the direct Higgs couplings,
it is natural to try and look for a more economical alternative.
This provides a motivation to study quark and lepton mixing
in addition to Higgs mixing, as a means of generating the desired
non-renormalisable operators. For example in Fig.\ref{Fig4}
we show a spaghetti diagram which can generate 32 mixing 
along the $U^c$ line.
Notice that we have added an intermediate quark $U^{(0)}$ with the same
quantum numbers under the gauge symmetry of the model as $U_3^c$. In
general, 
both will mix with $\bar{U}^{(0)}$ through a heavy mass term and we may always
rotate the definition of the fields such that one of the linear combinations
is massless. We identify this combination with the (conjugate) right handed
top superfield of the MSSM and the other with the heavy field involved in the
mass suppression of the spaghetti diagrams\footnote{Wherever two fields with
identical quantum numbers are present, one of which is a state of the MSSM,
we will assume that this mixing has already been accounted for. The states
labeled as MSSM states will thefore be defined as the massless linear
combination.}.
This means that vector quark and lepton states 
must be added, one chiral partner of which has the same $X$ charge as a MSSM
state. 
In Fig.\ref{Fig5} we show how 32
mixing can be generated by mixing along the $Q$ line.

Similar diagrams can be drawn for the other mixings,
and by consideration of such diagrams we can see that 
vector fields with $X$ charges from -4 to 4 
in integer steps are required. For example complete 
mixing along the $Q$ line requires:
\bea
  Q^{(-4)}, \bar{Q}^{(4)}, 
  Q^{(-3)}, \bar{Q}^{(3)}, 
 Q^{(-2)},  \bar{Q}^{(2)},
  Q^{(-1)},  \bar{Q}^{(1)},
 Q^{(0)}, \bar{Q}^{(0)},\nn \\
  Q^{(1)}, \bar{Q}^{(-1)}, 
 Q^{(2)},  \bar{Q}^{(-2)},
  Q^{(3)},  \bar{Q}^{(-3)},
 Q^{(4)}, \bar{Q}^{(-4)}.
\label{Qlist}
\eea
corresponding to 9 vector $Q+\bar{Q}$ representations.
Also, mixing along the $U^c$ line requires 
9 vector $U^c+\bar{U^c}$ representations;
mixing along the $D^c$ line requires 
9 vector $D^c+\bar{D^c}$ representations;
mixing along the $L$ line requires 
9 vector $L+\bar{L}$ representations; and
mixing along the $E^c$ line requires 
9 vector $E^c+\bar{E^c}$ representations.
In each case the $X$ charges run from -4 to +4 in integer steps
(plus the opposite charges for the conjugate states),
as in Eq.\ref{Qlist}.

Now that we have allowed quark and lepton mixing (along the
doublet and/or singlet lines) as well as Higgs mixing, many 
possibilities present themselves, corresponding to different
combinations of each type of mixing.
In the next section we shall use the constraint of string gauge
unification to help to discriminate between the different possibilities.
Here we shall make some general observations about the model
building.

Let us begin with the $U$ mass matrix. 
We can envisage a scenario in which we have the 9 $Q+\bar{Q}$ fields
listed in Eq.\ref{Qlist}, plus
9 vector $U^c+\bar{U^c}$ representations. In addition to these we also
have the three chiral families with the $X$ charges -4,1,0 as discussed above.
To be more general we must also consider additional Higgs
vector representations. We have seen that a possible Higgs mixing
scenario requires 10 $H_2+\bar{H_2}$ plus 10 $H_1+\bar{H_1}$
extra Higgs with $H$ charges from 8 to -2. However we may also wish
to consider Higgs which couple any vector quark field to any other vector
quark field, in which case the Higgs charges must range from 8 to -8.
If for the moment we ignore the MSSM fields, but include all of the
extra vector fields mentioned above then we have a $9\times 9$
matrix of Higgs couplings to quark fields:
\begin{eqnarray}
\begin{array}{rlllllllll}
\ \ &  U^{(-4)} &   U^{(-3)} &  U^{(-2)}  
&  U^{(-1)} &  U^{(0)} &   U^{(1)} &  U^{(2)} &  U^{(3)} &  U^{(4)}\\
Q^{(-4)}: &   H_{2}^{(8)} & H_{2}^{(7)} & H_{2}^{(6)}
& H_{2}^{(5)} & H_{2}^{(4)} & H_{2}^{(3)}
& H_{2}^{(2)} & H_{2}^{(1)} & H_{2}^{(0)}\\
Q^{(-3)}: &  H_{2}^{(7)} & H_{2}^{(6)} & H_{2}^{(5)} 
& H_{2}^{(4)} & H_{2}^{(3)} & H_{2}^{(2)} 
& H_{2}^{(1)} & H_{2}^{(0)} & H_{2}^{(-1)} \\
  Q^{(-2)}: &   H_{2}^{(6)} & H_{2}^{(5)} & H_{2}^{(4)} 
& H_{2}^{(3)} & H_{2}^{(2)} & H_{2}^{(1)} 
& H_{2}^{(0)} & H_{2}^{(-1)} & H_{2}^{(-2)} \\
  Q^{(-1)}: &  H_{2}^{(5)} & H_{2}^{(4)} & H_{2}^{(3)} 
& H_{2}^{(2)} & H_{2}^{(1)} & H_{2}^{(0)} 
& H_{2}^{(-1)} & H_{2}^{(-2)} & H_{2}^{(-3)}  \\
  Q^{(0)}: & H_{2}^{(4)} & H_{2}^{(3)} & H_{2}^{(2)} 
& H_{2}^{(1)} & H_{2}^{(0)} & H_{2}^{(-1)} 
& H_{2}^{(-2)} & H_{2}^{(-3)} & H_{2}^{(-4)} \\
  Q^{(1)}: &  H_{2}^{(3)} & H_{2}^{(2)} & H_{2}^{(1)} 
& H_{2}^{(0)} & H_{2}^{(-1)} & H_{2}^{(-2)} 
& H_{2}^{(-3)} & H_{2}^{(-4)} & H_{2}^{(-5)} \\
  Q^{(2)}: &  H_{2}^{(2)} & H_{2}^{(1)} & H_{2}^{(0)} 
& H_{2}^{(-1)} & H_{2}^{(-2)} & H_{2}^{(-3)} 
& H_{2}^{(-4)} & H_{2}^{(-5)} & H_{2}^{(-6)} \\
  Q^{(3)}: &  H_{2}^{(1)} & H_{2}^{(0)} & H_{2}^{(-1)} 
& H_{2}^{(-2)} & H_{2}^{(-3)} & H_{2}^{(-4)} 
& H_{2}^{(-5)} & H_{2}^{(-6)} & H_{2}^{(-7)} \\
  Q^{(4)}: &  H_{2}^{(0)} & H_{2}^{(-1)} & H_{2}^{(-2)} 
& H_{2}^{(-3)} & H_{2}^{(-4)} & H_{2}^{(-5)} 
& H_{2}^{(-6)} & H_{2}^{(-7)} & H_{2}^{(-8)}
\end{array}
\label{BigHiggsmatrix}
\end{eqnarray}
There is of course a similar matrix of couplings involving the
conjugate fields. Let us now introduce the three families of
MSSM chiral fields shown below into the matrix
in Eq.\ref{BigHiggsmatrix}:
\beq
  Q_1^{(-4)}, 
  Q_2^{(1)}, 
 Q_3^{(0)},  
  U_1^{(-4)}, 
  U_2^{(1)},
 U_3^{(0)}, 
\eeq
where the family index is indicated by a subscript and the 
$X$ charge is indicated by a superscript.
The matrix now becomes a $12 \times 12$ matrix,
and the Higgs content stays the same.
We can now consider all possible ways in which mass mixing
between the MSSM quarks can occur.
In general for ij mixing we require a spaghetti diagram with
the external lines consisting of $Q_i$, $U_j$, $H_2^{(0)}$.
In addition we are allowed to hang any amount of
$\theta$ and $\bar{\theta}$ spaghetti along
any of the three lines $Q,U,H_2$ in order to achieve the mixing
where the minimum amount of spaghetti
corresponds to the leading non-renormalisable operator.
To take a trivial example the 33 operator $Q_3U_3H_2^{(0)}$
is achieved directly at tree level without any
spaghetti. At the other extreme the 11 operator
$Q_1^{(-4)}U_1^{(-4)}H_2^{(0)}$ is clearly forbidden at tree level
by the $X$ symmetry, with the allowed non-renormalisable
operator being $Q_1^{(-4)}U_1^{(-4)}H_2^{(0)}(\theta^{(1)})^8$.
The required 8 pieces of $\theta$ spaghetti can be hung
along any of the three lines $Q,U,H_2$ depending on the 
vector fields and charges which are assumed.
For example in the Higgs mixing scenario of IR there is a
tree level Higgs coupling $Q_1^{(-4)}U_1^{(-4)}H_2^{(8)}$
and the MSSM Higgs $H_2^{(0)}$ is achieved by stepping down the
Higgs charge along the first row of the matrix
in Eq.\ref{BigHiggsmatrix} with the Higgs charge decreasing by one
unit after each $\theta$ field insertion:
\beq
H_2^{(8)}\stackrel{\theta^{(1)}}{\ra }H_2^{(7)}
         \stackrel{\theta^{(1)}}{\ra }H_2^{(6)}
\stackrel{\theta^{(1)}}{\ra }H_2^{(5)}
\stackrel{\theta^{(1)}}{\ra }H_2^{(4)}
\stackrel{\theta^{(1)}}{\ra }H_2^{(3)}
\stackrel{\theta^{(1)}}{\ra }H_2^{(2)}
\stackrel{\theta^{(1)}}{\ra }H_2^{(1)}
\stackrel{\theta^{(1)}}{\ra }H_2^{(0)}
\eeq
with all the spaghetti mixing along the Higgs line.
With the additional vector $Q$ and $U^c$ fields
considered above there are alternative ways in which 
the spaghetti mixing can take place.
For example we could begin from the non-renormalisable operator
$Q_1^{(-4)}U^{(4)}H_2^{(0)}$ and step down the the $U$ line
to reach the desired $U_1^{(-4)}$ field:
\beq
U^{(4)}\stackrel{\bar{\theta}_U^{(1)}}{\ra }U^{(3)}
\stackrel{\bar{\theta}_U^{(1)}}{\ra }U^{(2)}
\stackrel{\bar{\theta}_U^{(1)}}{\ra }U^{(1)}
\stackrel{\bar{\theta}_U^{(1)}}{\ra }U^{(0)}
\stackrel{\bar{\theta}_U^{(1)}}{\ra }U^{(-1)}
\stackrel{\bar{\theta}_U^{(1)}}{\ra }U^{(-2)}
\stackrel{\bar{\theta}_U^{(1)}}{\ra }U^{(-3)}
\stackrel{\bar{\theta}_U^{(1)}}{\ra }U_1^{(-4)}
\eeq
The resulting spaghetti diagram now has all the mixing along the $U$ line,
but is of the same order as the previous diagram.
We could repeat this starting instead from the 
non-renormalisable operator
$Q^{(4)}U_1^{(-4)}H_2^{(0)}$ and step down the the $Q$ line
to reach the desired $Q_1^{(-4)}$ field:
\beq
Q^{(4)}\stackrel{\bar{\theta}_Q^{(1)}}{\ra }Q^{(3)}
\stackrel{\bar{\theta}_Q^{(1)}}{\ra }Q^{(2)}
\stackrel{\bar{\theta}_Q^{(1)}}{\ra }Q^{(1)}
\stackrel{\bar{\theta}_Q^{(1)}}{\ra }Q^{(0)}
\stackrel{\bar{\theta}_Q^{(1)}}{\ra }Q^{(-1)}
\stackrel{\bar{\theta}_Q^{(1)}}{\ra }Q^{(-2)}
\stackrel{\bar{\theta}_Q^{(1)}}{\ra }Q^{(-3)}
\stackrel{\bar{\theta}_Q^{(1)}}{\ra }Q_1^{(-4)}
\eeq
The resulting spaghetti diagram now has all the mixing along the $Q$ line,
but is of the same order as the previous diagram.
There are of course many other possibilities which involve
a combination of the three types of mixing, and all these possibilities
will lead to non-renormalisable operators of the same order.
For example suppose we again wish to generate the 11 operator
$Q_1^{(-4)}U_1^{(-4)}H_2^{(0)}(\theta^{(1)})^8$ starting from
the tree level operator $Q^{(-2)}U^{(-2)}H_2^{(4)}$. Now in order to
achieve this we must have all three types of mixing simultaneously:
\beq
H_2^{(4)}
\stackrel{\theta^{(1)}}{\ra }H_2^{(3)}
\stackrel{\theta^{(1)}}{\ra }H_2^{(2)}
\stackrel{\theta^{(1)}}{\ra }H_2^{(1)}
\stackrel{\theta^{(1)}}{\ra }H_2^{(0)}
\eeq
\beq
U^{(-2)}
\stackrel{\bar{\theta}_U^{(1)}}{\ra }U^{(-3)}
\stackrel{\bar{\theta}_U^{(1)}}{\ra }U_1^{(-4)}
\eeq
\beq
Q^{(-2)}
\stackrel{\bar{\theta}_Q^{(1)}}{\ra }Q^{(-3)}
\stackrel{\bar{\theta}_Q^{(1)}}{\ra }Q_1^{(-4)}
\eeq
Again the 11 operator is eighth order, but now
there are four pieces of spaghetti from the Higgs line,
two from the $U$ line and two from the $Q$ 
line. There are clearly many other possible ways of achieving
11 $U$ mixing. The discussion of the other $U$ mixings is similar.
Finally the discussion of the $D$ and $E$ mixing matrices follows
a similar pattern.
The most general model clearly involves 9 vector copies
of each of 
$(Q+\bar{Q}),(U+\bar{U}),(D+\bar{D}),(L+\bar{L}),(E+\bar{E})$,
where we denote the number of vector copies
as $n_Q,n_U,n_D,n_L,n_E$ respectively,
plus 16 vector copies of each of the Higgs fields 
$(H_2+\bar{H_2}),(H_1+\bar{H_1})$, where we denote the number of vector 
copies as
$n_{H_1},n_{H_2}$, respectively.

This is clearly not the most
economical model. For example the Ross\cite{GG} model is based on
no extra vector quarks and leptons and 
10 vector copies of each of
$(H_2+\bar{H_2}),(H_1+\bar{H_1})$. In the next section we shall impose the 
constraint of string gauge unification in order to try to determine a more
economical model. However it is clear from the discussion of this section,
that if too few extra vector states are allowed, then the required mass mixing
will
not be achievable. Therefore we seek the minimal model
which is consistent with the constraints of spaghetti mixing discussed here,
and string gauge unification. Since mixing can be achieved 
by a combination of mixing along the three different lines
of the spaghetti diagram, in the next section we shall impose the
following conservative lower limits on the minimal numbers of 
vector copies of fields.
From $U$ mixing we require:
\beq
n_Q + n_2 + n_U  \geq  8. \label{up}
\eeq
From $D$ mixing we require:
\beq
n_Q + n_2 + n_D \geq  8. \label{dowb}
\eeq
From $E$ mixing we require:
\beq
n_2 + n_E  \geq  8. \label{lepton}
\eeq
where for convenience we have defined the total number of doublets as
\beq
n_2 \equiv n_{H_1}+n_{H_2}+n_L.
\eeq
In addition we shall allow for 
exotic superfields known as ``sextons'' $S$ which 
are colour triplets, electroweak singlets and have hypercharge
$1/6$. The sextons occur
together with their vector conjugate 
$(S+\bar{S})$ and we denote the number of vector copies of sextons as $n_S$. 
These superfield representations are present in the massless spectrum of some
string models~\protect\cite{faraggi}.

\section{String gauge unification analysis}

We now describe the numerical constraints placed upon the models to ensure
that they 
provide agreement with the low energy data, string scale gauge unification and
are compatible with models of fermion mass and mixing.
Since we are hoping to eventually embed the model into a free-fermionic
Kac-Moody level 1 string model, we must make sure that the gauge couplings
obey the constraint of
gauge unification at the string scale as in Eq.\ref{stringscale}. 
Another constraint comes from the agreement with the empirically determined
values of the gauge couplings at low energy scales~\cite{pdb}. The data used
is
\bea
\alpha(M_Z)^{-1}&=&127.9 0 \pm 0.09 \nn \\
\sin^2 \theta_w &=& 0.2315 \pm 0.0002\nn \\
\alpha_S (M_Z) &=& 0.118 \pm 0.003, \label{dataused}
\eea
where the numbers quoted are those derived in the
$\o.{MS}$ renormalisation scheme from experiments. In what follows, we
shall assume
central values for $\alpha(M_Z)^{-1}, \sin^2 \theta_W$ because their errors
are comparatively small.
The third constraint comes from the fact that we are expecting to use the
intermediate matter as the heavy fields in a family U(1)$_X$ model of
fermion masses. As previously demonstrated~\cite{IR}, these models require
$M_I$ to be within a couple of orders of magnitude of $M_X$ to make the
GS gauge anomaly cancellation mechanism work consistently. After some
other filtering of models, as described below, the condition imposed on any
successful model will be
\beq
M_I / M_X \geq 1 / 40. \label{scales}
\eeq
Previous models of U(1)$_X$ family symmetry require a GUT type
normalisation of $Y$ from the anomaly cancellation conditions.
Therefore, our condition upon the unification of gauge couplings
will be
\beq
g_1 (M_X)\equiv \sqrt{\frac{5}{3}} \frac{Y(M_X)}{2}  = g_2(M_X) = g_3(M_X)
\label{unprecise},
\eeq
corresponding to a Kac-Moody level 1 string model with $k_1 = 5/3$.
Finally, bearing in mind the long-term view of requiring the intermediate
sector to
mediate masses and mixings to all of the SM fermions, we impose
Eqs.\ref{up}-\ref{lepton}
We then search through all of the models satisfying Eqs.\ref{up}-\ref{lepton}
for
$n_Q$, $n_2$, $n_U$, $n_D$, $n_E$ $\leq 10$ in order to find the
models with less field content. 

We now describe the systematic procedure to determine the models that pass the
constraints\footnote{In fact, we will allow the prediction of $\alpha_S(M_Z)$
to be within $2\sigma$ of the value quoted in Eq.\protect\ref{dataused}.}
given by Eqs.\ref{up}-\ref{lepton},\ref{dataused}-\ref{unprecise}. 
Because of computer time constraints, we were not able to determine the
predictions to two-loop order of every model
satisfying Eqs.\ref{up}-\ref{lepton}. The following procedure was
therefore adopted: the predictions for every different choice of intermediate
field content were determined to one loop
order as in~\cite{MSSM+X} for $k_1=5/3$. 
If the one-loop predictions did not satisfy certain constraints to be
described
shortly, the models were discarded.
For any models passing the previous ``cut'', the predictions were obtained at
two-loop order.

The first one-loop filtering procedure was as follows: once a model has been
selected by a a particular choice of intermediate matter, a guess
$({M_X}')$ of the 
string unification scale was made. The condition
$\alpha_1({M_X}')=\alpha_2({M_X}')$ 
yields a value of $M_I$ consistent with unification at $M_X$ by solving the
one-loop RGEs for $\alpha_1$ and $\alpha_2$ in the $\o.{MS}$ scheme to
obtain
\bea
\ln M_I &=& \left( 2 \pi (\alpha_2^{-1}(M_Z) - \alpha_1^{-1}(M_Z)) -
\frac{77}{10} \ln M_Z
+ 
\frac{13}{30} \ln m_t + 
\frac{5}{3} \ln M_{SUSY} \r. \nn \\ &&\l.
-(3 n_Q +
n_2 -
\frac{56}{10} - \frac{6}{10} Y_T) \ln {M_X}' \right) / (Y_T \frac{3}{5} - 3n_Q
- n_2),
\eea
where $Y_T\equiv \sum_i (Y_i/2)^2$, $i$ runs over all of the intermediate
states and $Y_i$ denotes the hypercharge of the intermediate state $i$.
Throughout the numerical analysis, the
entire SUSY spectrum was assumed to be at an effective scale equal to
$M_{SUSY}$. Any threshold effects were taken into account by a step function
approximation. 
With $M_I$ and ${M_X}'$ values consistent with gauge unification at a scale
${M_X}'$, we could calculate $M_X$
consistent with string scale unification in Eq.\ref{stringscale} by finding
\bea
\alpha_2^{-1}(M_X) &=& \alpha_2^{-1} (M_Z) + \frac{25}{12 \pi} \ln
\frac{m_t}{M_Z} + \frac{19}{12 \pi} \ln \frac{M_{SUSY}}{m_t} - \frac{1}{2 \pi}
\ln \frac{M_I}{M_{SUSY}} 
\nn \\
&&- \frac{1 + 3 n_Q + n_2}{2\pi}\ln\frac{M_X}{M_I} 
\eea
and substituting it into Eq.\ref{stringscale}. To obtain values of $M_X, M_I$
consistent both
with $\alpha_1(M_X)=\alpha_2(M_X)$ and Eq.\ref{stringscale}, we now
substitute ${M_X}'$
with $M_X$ and iterate the above procedure until $M_X={M_X}'$ is satisfied to
some required accuracy. This yields a prediction for $\alpha_3(M_Z)$ by
using $\alpha_3(M_X)\equiv \alpha_G$ where $\alpha_G$ is the string scale
unified gauge structure constant. $\alpha_3$ is then run to low energies
using the one loop RGEs,
\bea
\alpha_3^{-1}(M_Z) &=& \alpha_G(M_X) + \frac{23}{6 \pi} \ln M_Z - \frac{1}{3
\pi} \ln m_t - \frac{2}{\pi} \ln M_{SUSY} - 
\nn \\ && \frac{2 n_Q + n_U + n_D}{2 \pi} \ln
M_I - \frac{3 - 2 n_Q - n_U  - n_D}{2 \pi} \ln M_X.
\eea
We have set $M_{SUSY}=m_t$, which\footnote{$m_t$ denotes the running top mass
in the
$\o.{DR}$ renormalisation scheme.} we take to be 166 GeV, corresponding
to a top quark pole mass of 180 GeV for central values of $\alpha_S(M_Z)$, as
in ref.~\cite{pdb}. We will return later to the effect of the empirical errors
upon the inputs.
We now require each model to pass the cuts
$\alpha_S(M_Z) 
\leq 0.124$ and $M_I / M_X \geq 1/100$ to be worthy of the two-loop
analysis. Note that these constraints are purposefully less severe than the
ones in
Eqs.\ref{dataused},\ref{scales} because we do not want to discard models in
which the imprecise one-loop predictions do not pass
the more restrictive constraints, but in which the two-loop predictions pass.

Having attained a list of all models that passed the initial cuts, the
two-loop predictions were then attained. At the two-loop level, the third
family Yukawa couplings all effect the running of the gauge couplings and
therefore the predictions of gauge unification. To a good
approximation, the other
Yukawa couplings of the MSSM have a negligible effect upon the running. 
As a starting point we must then obtain the values of these couplings at a
particular scale, for a chosen value of $\tan \beta$ and $\alpha_S(M_Z)$.
Once we have selected these two parameters, we may determine the renormalised
masses $m_{t,b,\tau}(m_t)$ of the
top, bottom
and tau particles at the renormalisation scale $m_t$, by
running the three loop QCD $\otimes$ one loop QED RGEs through quark
thresholds
between $m_\tau$ and $M_{SUSY}$ in the $\o.{MS}$
scheme~\cite{3loopQCD}. $\alpha_S(\mu <M_Z)$ is actually run using a state of 
the art four loop QCD beta function~\cite{4loopQCD}. 
In fact, the four loop contribution only
changes the predictions by a few parts per thousand.
This fact has a limited significance because the coefficient functions
required to extract $\alpha_S(M_Z)$ from data are only known to at most three
loops.
The three gauge couplings are also run between $M_Z$
and $m_t$ in the $\o.{MS}$ scheme, assuming the particle spectrum of
The Standard Model without the Higgs or top quark. The third-family
$\o.{MS}$ Yukawa 
couplings may then be determined by~\cite{btauNMSSM}
\bea
\lambda_t (m_t) &=& \frac{m_t(m_t)\sqrt{2}}{v \sin \beta} \nn \\
\lambda_{b,\tau} (m_t) &=& \frac{m_{b,\tau}(m_t)\sqrt{2}}{v \cos \beta},
\label{matchyuk}
\eea
where $v=246.22$ GeV is the scale parameter of electroweak symmetry breaking.
Above $m_t$, we wish to use the RGEs for the MSSM contained in
Appendix~1.
However these are in the $\o.{DR}$ scheme and so at $m_t$ we match all
of the quantities obtained in the $\o.{MS}$ scheme to the
$\o.{DR}$ scheme. A guess for the intermediate scale $M_I$ is chosen and
the gauge couplings and third family Yukawa couplings are run to this scale.
Above $M_I$, the effect of the intermediate matter is felt and the RGEs change
as prescribed in Appendix~1. The couplings are run up in scale until either it
becomes non-perturbative (which we take to be greater than 4) or
until $g_1(\mu)=g_2(\mu)$. In the first case the value of $M_I$ chosen is
abandoned and in the second a prediction for $\alpha_3(m_t)$ is obtained by
using the gauge unification condition. This is implemented by setting
$g_3(\mu)=g_2(\mu)$ and then running all of the couplings down to $m_t$ taking
the intermediate matter into account and integrating it out of the effective
theory at $M_I$. We may now iterate the above
procedure using the previous predicted value of $g_3(m_t)$ as an input each
time until $\alpha_3(m_t)$ converges and we have a value consistent with
gauge unification with the intermediate matter at the guess value of $M_I$.
The above procedure is then repeated for different values of $M_I$ until a
value is found in which the gauge couplings and unification scale satisfy
Eq.\ref{stringscale}, i.e.\ the constraint of string scale gauge unification.
It is a simple matter to re-convert $\alpha_3(m_t)$ back into the
$\o.{MS}$ scheme and run back down to determine $\alpha_S(M_Z)$.
Thus, for a given $\tan \beta$, we now have the predictions $M_I,
\alpha_S(M_Z)$ that come from the assumption of string scale gauge
unification. The conditions in Eqs.\ref{dataused},\ref{scales} are then
employed to remove any models which do not agree with the data or fit into 
the type of
models of fermion masses being considered.

Tables~\ref{tab:mainresults},\ref{tab:res2} display the results
of the algorithm described above for
$\tan \beta=43,5$ respectively. The constraints we
impose upon the models are so tight that out of the tens of thousand models
examined, only a few models pass the constraints in each case.
Note that for $\tan \beta=43$, only two of these give $\alpha_S(M_Z)$
predictions
within $1\sigma$ of the central value. 
\begin{table}
\begin{center}
\begin{tabular}{|cccccccc|} \hline
$n_2$ & $n_Q$ & $n_U$ & $n_D$ & $n_E$ & $\alpha_3(M_Z)$ & $M_X/10^{18}$ GeV &
$M_I/10^{17}$ GeV \\
\hline
  0&  9&  4& 10&  8&0.1173&0.5763&0.3387\\
  0& 10&  5& 10&  8&0.1236&0.5457&0.5416\\
  1&  9&  5& 10&  7&0.1172&0.5925&0.3428\\
  1& 10&  6& 10&  7&0.1235&0.5565&0.5466\\
  2& 10&  7& 10&  6&0.1234&0.5681&0.5520\\
  3& 10&  8& 10&  5&0.1233&0.5808&0.5579\\
  4& 10&  9& 10&  4&0.1232&0.5946&0.5644\\
  5& 10& 10& 10&  3&0.1231&0.6098&0.5716\\ \hline
\end{tabular}
\end{center}
\caption{Predictions of models that successfully unify the gauge couplings at
$M_X$ and provide enough intermediate matter to build a model of fermion
masses for $\tan \beta=43$. }
\label{tab:mainresults}
\end{table}
\begin{table}
\begin{center}
\begin{tabular}{|cccccccc|} \hline
$n_2$ & $n_Q$ & $n_U$ & $n_D$ & $n_E$ & $\alpha_3(M_Z)$ & $M_X/10^{18}$ GeV &
$M_I/10^{17}$ GeV \\
\hline
  0&  9&  4& 10&  8&0.1198&0.5700&0.3696\\
  1 & 9 & 5 &10&  7&0.1196&0.5850&0.3737\\ \hline
\end{tabular}
\end{center}
\caption{Predictions of models that successfully unify the gauge couplings at
$M_X$ and provide enough intermediate matter to build a model of fermion
masses for $\tan \beta=5$. }
\label{tab:res2}
\end{table}
The minimum number of extra vector multiplets added is 32.
Varying $m_t^{phys},\sin^2 \theta_w$ between their $1 \sigma$ limits can make a
difference to $\alpha_S(M_Z)$ predictions of $\sim \pm 0.002$. The $M_I/M_X$
prediction is hardly affected by the change in the input parameters. Varying
$\alpha_1(M_Z)$ within its $1 \sigma$ limits makes only a negligible change to
the predictions.
We note that the greatest uncertainty in our two-loop calculation is likely to
be that due to threshold effects. Until now we have assumed that all of the
intermediate matter lies at one scale $M_I$. While this is in some sense the
simplest scheme, in general there could be some splittings between the
different types of additional matter. One may naively expect these to not span
more than one order of magnitude, but even given this constraint there could
be significant errors due to the non-degeneracy. Non-degeneracy of the
superpartner spectrum could also cause errors in the predictions.
If we allow the presence of sexton fields, there are some
additional possible models that the algorithm just described will not
find. These are models with equal one-loop beta functions above $M_I$.
In this case, the one-loop algorithm fails because the solution to
$g_1(\mu)=g_2(\mu)=g_3(\mu)$ is not unique. Above $M_I$,
the gauge couplings have slopes that differ by small two-loop effects.
It is a simple matter to demonstrate that the models
\bea
n_Q&=&0,\ n_S=24,\ n_2-n_D=20; \nn \\
n_Q&=&1,\ n_S=16,\ n_2-n_D=11; \nn \\
n_Q&=&2,\ n_S=8,\ n_2-n_D=2; \nn \\
n_Q&=&3,\ n_S=0,\ n_2-n_D=7; \label{equal}
\eea
have the property of equal one-loop beta functions above $M_I$.
\begin{table}
\begin{center}
\begin{tabular}{|ccccccccc|} \hline
$n_2$ & $n_Q$ & $n_U$ & $n_D$ & $n_E$ & $n_S$ & $\alpha_3(M_Z)$ &
$M_X/10^{18}$ GeV & $M_I/10^{17}$ GeV \\ \hline
 20&  0&  0&  0&  0&24&0.1218&0.4979&0.2906\\
 21 & 0 & 0 & 1 & 0&24&0.1218&0.5069&0.2940\\
 11 & 1&  0&  0&  0&16&0.1226&0.4529&0.2788\\
 12&  1&  0&  1&  0&16&0.1225&0.4593&0.2815\\ \hline
\end{tabular}
\end{center}
\caption{Predictions of models that successfully unify the gauge couplings at
$M_X$ and provide enough intermediate matter to build a model of fermion
masses for $\tan \beta=43$. The models shown here belong to the special class
of models that possess equal one-loop beta functions.}
\label{tab:eq}
\end{table}
Some of these models were investigated with an accurate version of the
two-loop algorithm.
Table~\ref{tab:eq} displays the predictions of a subset of the successful
models in
Eq.\ref{equal}. 

\section{Infra-Red Fixed Points}

We now turn to the question of infra-red fixed points (IRFPs)
of the dimensionless Yukawa couplings
for the class of models which are consistent with the generation
of acceptable textures via spaghetti mixing, and string gauge unification.
Our discussion follows that of the IRFPs for the Ross model of Higgs
mixing \cite{GG}. The basic idea behind this approach is 
that when the large Yukawa couplings between heavy fields and Higgs' are
renormalised, their low energy values may not be sensitive to the high energy
ones, and so a rough prediction of the value may be made. This is the
situation around an infra-red stable fixed point, where combinations of
dimensionless couplings in the theory stop changing with renormalisation.
Such fixed points are very
welcome in this approach since our textures result from 
a large number of unknown Yukawa couplings, which would otherwise
render this approach quite unpredictive.

There is another reason for looking for infra-red fixed points in the models
discussed here.
The presence of a gauged family symmetry such as $U(1)_X$ is in principle
quite dangerous since its presence
can lead to large off-diagonal squark and slepton
masses which can mediate flavour-changing processes at low energy.
In particular the $D$ term associated with $U(1)_X$
is in general only approximately flat due to lifting by soft supersymmetry
breaking
terms, and this can lead to family-dependent squark and slepton masses  
with unacceptably large mass splittings. 
This is a generic problem
of any model with a gauged family symmetry, however the $U(1)_X$ symmetry
here is non-asymptotically free with a large beta function so that its   
gauge coupling rapidly becomes very small below the string scale,
leading to small $X$ gaugino masses.

It has been suggested \cite{RLfixed} that the possible infra-red structure
of the theory could help by relating the soft scalar masses to the small
gaugino masses, thereby making them naturally smaller than the
squark and slepton masses,
or by enforcing $<\theta>=<\bar{\theta}>$ as an infra-red
fixed point of the theory. 
We refer the reader to ref.\cite{RLfixed} for more
details. 
Here we shall only focus on the IRFPs of the
dimensionless Yukawa couplings however.

\subsection{The Top Quark Yukawa Coupling}

\begin{figure}
\begin{center}
\leavevmode   
\hbox{\epsfxsize=3.5in
\epsfysize=3.5in
\epsffile{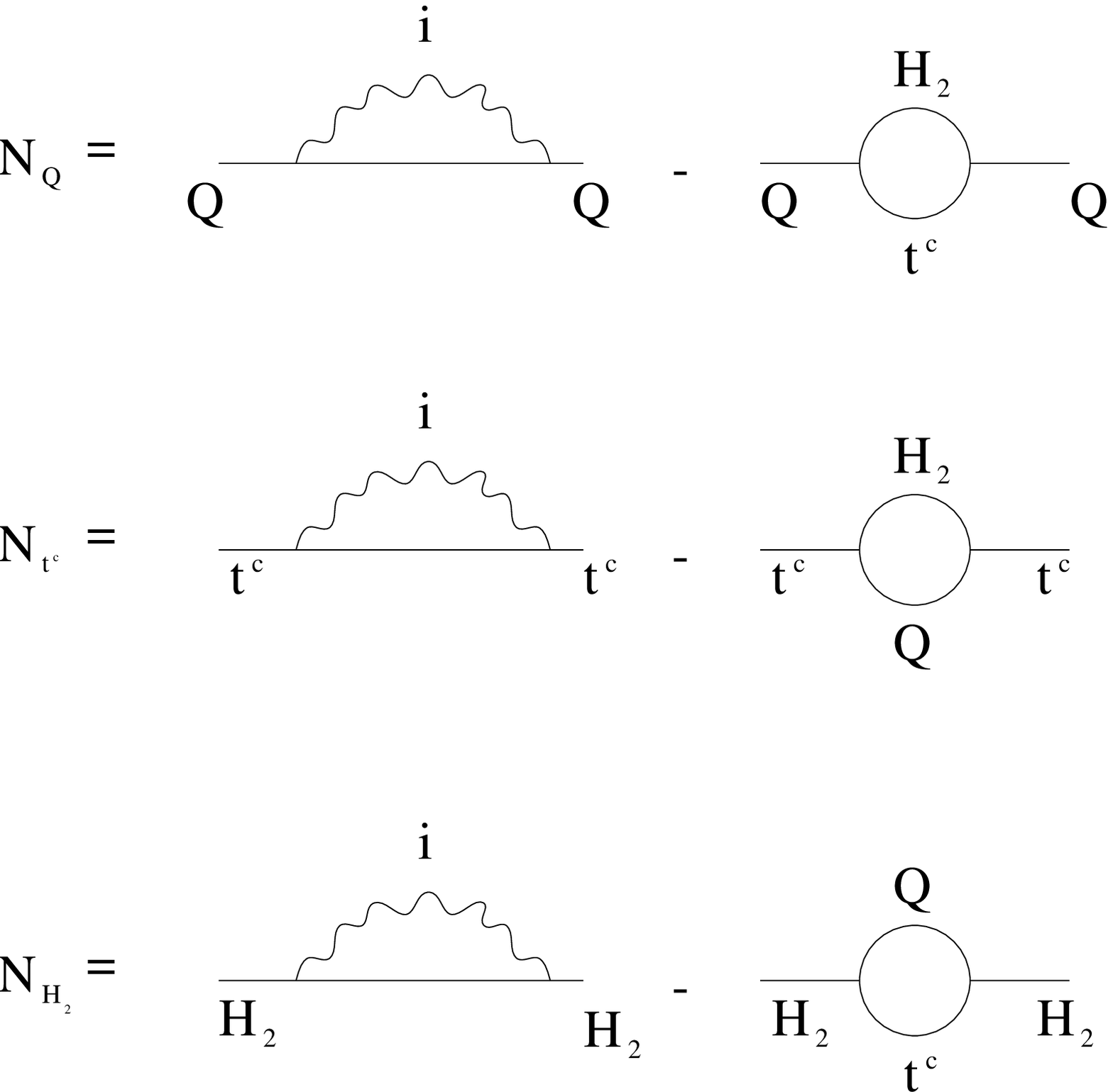}}
\end{center}
\caption{Contributions to the wavefunction renormalisation
of the top quark Yukawa coupling.}
\label{Fig6}
\end{figure}

The first step in finding the IRFPs of the theory is to construct
the RGEs of the dimensionless Yukawa
couplings of the theory. In supersymmetric theories this task is made
quite simple, at least at the one loop level, by the observation that
only the wavefunction diagrams contribute to the RGEs.
The vertex contributions vanish due to the non-renormalisation theorem.
This allows the RGEs to be constructed in a very straightforward manner.
To take a simple example, consider a toy theory
which only involves the top quark Yukawa coupling in the superpotential:
\beq
W=hQt^cH_2.
\eeq
Defining the Yukawa and three gauge coupling parameters as 
\beq
Y^h=\frac{h^2}{16\pi^2}, \  \tilde{\alpha_i}=\frac{g_i^2}{16\pi^2},
\eeq
and the scale variable as
\beq
t=-\ln (\mu^2)
\eeq
we can write the RGE for the top quark Yukawa coupling as
\beq
\frac{dY^h}{dt} = Y^h(N_Q+N_{t^c}+N_{H_2})
\eeq 
where $N_i$ are the wavefunction renormalisation contributions
from each of the three legs of the vertex.
In the toy model the wavefunction diagrams are shown in Fig.\ref{Fig6}.

The wavefunction renormalisation contributions are explicitly,
\bea
N_Q     & = & \sum_i 2C_{2i}(Q)\tilde{\alpha_i} -Y^h \nn \\
N_{t^c} & = & \sum_i 2C_{2i}(t^c)\tilde{\alpha_i} -2Y^h \nn \\
N_{H_2} & = & \sum_i 2C_{2i}(H_2)\tilde{\alpha_i} -3Y^h
\eea
where $2C_{2i}(R)$ is the quadratic Casimir of the representation
$R$ under the $i-th$ gauge group factor of the MSSM,
arising from the gauge boson exchange corrections,
and the multiplicity factors in front of the $Y^h$ terms
are due to doublet and colour counting for the particles
going round the loop.
Thus the RGE is explicitly
\beq
\frac{dY^h}{dt} = Y^h(\sum_{i,R} 2C_{2i}(R)\tilde{\alpha_i} -6Y^h)
\eeq 
Now let us assume that in our toy model
all three gauge couplings were equal, and
all three gauge beta functions were equal (quite unrealistic
for the low energy Yukawa coupling, but typical of the situation
near the string scale).
Then the RGE may be written as
\beq
\frac{dY^h}{dt} = Y^h(r\tilde{\alpha} -6Y^h)
\eeq 
where we have defined 
\beq
r\equiv \sum_{i,R} 2C_{2i}(R)
\eeq
where $R$ runs over all fields involved in the coupling $Y^h$. We have written
the three equal gauge couplings as
$\tilde{\alpha}$.
We now write the one loop gauge running as
\beq
\frac{\partial \tilde{\alpha}}{\partial t}=-b{\tilde{\alpha}}^2 \label{Bfns}
\eeq
and define the ratio of Yukawa to gauge coupling as
\beq
R^h\equiv \frac{Y^h}{\tilde{\alpha}} \label{ratio}
\eeq
Then the RGE for this ratio is:
\beq
\frac{dR^h}{dt} = {\tilde{\alpha}} R^h(r +b -6R^h)
\eeq 
and the Pendleton-Ross fixed point is given by
$\frac{dR^h}{dt}=0$. This condition can be achieved by
\beq
r+b-6R^h=0.
\eeq
In terms of wavefunction renormalisation parameters,
the RGE can be expressed as:
\beq
\frac{dR^h}{dt} =  R^h(N_Q+N_{t^c}+N_{H_2}+{\tilde{\alpha}}b).
\eeq 
The fixed point condition can be expressed as\footnote{It is worth noting that
an
alternative fixed point has recently been proposed by Jack and
Jones \cite{JJ}, in which $\frac{dR^h}{dt}=0$ is achieved by
the more stringent conditions:$N_Q=N_{t^c}=N_{H_2}=
-\frac{1}{3}{\tilde{\alpha}}b$. These conditions may be expressed
in a more general way, and are valid to two loops.
However they are not satisfied for our simple toy model,
and they will not be of use for our more general model.}:
\beq
N_Q+N_{t^c}+N_{H_2}+{\tilde{\alpha}}b=0.
\eeq

\subsection{The General Model}

Ross \cite{GG} applied the above techniques to find the
IRFPs of the Higgs mixing model. We wish to extend the discussion
to the more general situation of $n_Q,n_U,n_D,n_L,n_E$
copies of the vector representations
\beq
(Q^{(x)}+\bar{Q}^{(-x)}),\ (U^{c(y)}+\bar{U}^{c(-y)}),
\ (D^{c(y)}+\bar{D}^{c(-y)}),\ (L^{(x)}+\bar{L}^{(-x)}),
\ (E^{(y)}+\bar{E}^{c(-y)})
\eeq
plus 
$n_{H_1},n_{H_2}$
copies of the vector representations
\beq
(H^{(z)}_{2}+\bar{H}^{(-z)}_{2}),\ (H^{(z)}_{1}+\bar{H}^{(-z)}_{1})
\eeq
where we have labeled the
vector fields by their $X$ charges $x,y,z$.
For the special case of $z=0$ the Higgs are identified
with the two MSSM Higgs doublets,
\beq
H^{(0)}_1, \ H^{(0)}_2
\eeq
and do not have vector conjugates.
The above fields are in addition to the three
chiral families of quarks and leptons which we label as:
\beq
Q_i,\ U^c_j, \ D^c_j, \ L_i, \ E^c_j
\eeq
where we label these fields by the family index subscript $i,j=1,\cdots 3$,
but do not label their $X$ charges (for $i,j=1,2,3$ the $X$ charges
are -4,1,0, respectively, as discussed earlier).
We introduce the $X$ charge breaking singlet Higgs fields
$(\theta_Q+\bar{\theta}_Q),(\theta_U+\bar{\theta}_U),
(\theta_D+\bar{\theta}_D),(\theta_L+\bar{\theta}_L),
(\theta_E+\bar{\theta}_E)$ plus
$(\theta_2+\bar{\theta}_2),(\theta_1+\bar{\theta}_1)$
which change the $X$ charge of the particular field
by 1 or -1. 
We also introduce MSSM singlet Higgs with $X=0$:
$\Phi_Q,\Phi_U,\Phi_D,\Phi_L,\Phi_E$ plus $\Phi_2,\Phi_1$
whose VEVs are responsible for the heavy vector masses at a common scale
$M_I$. 

The most general model is then defined by the gauge group,
\beq
SU(3)_C\times SU(2)_L \times U(1)_Y \times U(1)_X
\eeq
with the superpotential 
involving the chiral quarks and leptons 
containing the trilinear terms
\bea
W_1 & = & 
\sum_{i,j}h_{ij}Q_iU^{c}_jH^{(z)}_{2}  
+  \sum_{i,j}k_{ij}Q_iD^c_jH^{(z)}_{1}
 +  \sum_{i,j}l_{ij}L_iE^c_jH^{(z)}_{1}
\nn \\
& + & \sum_{i,y}h_{i(y)}Q_iU^{c(y)}H^{(z)}_{2}  
+  \sum_{i,y}k_{i(y)}Q_iD^{c(y)}H^{(z)}_{1}
 +  \sum_{i,y}l_{i(y)}L_iE^{c(y)}H^{(z)}_{1}
\nn \\
& + & \sum_{x,j}h_{(x)j}Q^{(x)}U^{c}_jH^{(z)}_{2}  
+  \sum_{x,j}k_{(x)j}Q^{(x)}D^c_jH^{(z)}_{1}
 +  \sum_{x,j}l_{(x)j}L^{(x)}E^c_jH^{(z)}_{1}
\nn \\
& + & \sum_i s_{Q_i}\theta_{Q}Q_i\bar{Q}^{(-x-1)}
+ \sum_j s_{U_j}\theta_{U}U^{c}_j\bar{U}^{c(-y-1)}
+ \sum_j s_{D_j}\theta_{D}D^{c}_j\bar{D}^{c(-y-1)}
\nn \\ 
& + &\sum_i s_{L_i}\theta_{L}L_i\bar{L}^{(-x-1)}
+  \sum_j s_{E_j}\theta_{E}E^{c}_j\bar{E}^{c(-y-1)}
\nn \\ 
& + & \sum_i \bar{s}_{Q_i}\bar{\theta}_{Q}Q_i\bar{Q}^{(-x+1)}
+ \sum_j \bar{s}_{U_j}\bar{\theta}_{U}U^{c}_j\bar{U}^{c(-y+1)}
+ \sum_j \bar{s}_{D_j}\bar{\theta}_{D}D^{c}_j\bar{D}^{c(-y+1)}
\nn \\ 
& + &\sum_i \bar{s}_{L_i}\bar{\theta}_{L}L_i\bar{L}^{(-x+1)}
+  \sum_j \bar{s}_{E_j}\bar{\theta}_{E}E^{c}_j\bar{E}^{c(-y+1)}. \label{frog}
\eea
In Eqs.\ref{frog}-\ref{bigsuperpotential}, it is to be understood that the $X$
charges of the
fields in 
each coupling must add to zero and that this decides the superscripts that are
not summed over. This is true for all superpotentials and wave-function
renormalisations listed in this paper.
We neglect some terms in the superpotential that are not banned by the
symmetries we have listed so far. Some of these are undesirable in terms of
reproducing the correct phenomenology, and so we appeal to the extra U(1)
symmetries that tend to come with string-derived models to ban these terms.
The remaining terms in the superpotential 
involving the extra vector states and Higgs are:
\bea
W_2 & = &  \sum_{x,y}h_{(xy)}Q^{(x)}U^{c(y)}H^{(z)}_{2}  
+  \sum_{x,y}k_{(xy)}Q^{(x)}D^{c(y)}H^{(z)}_{1}
\nn \\
& + & \sum_{x,y}l_{(xy)}L^{(x)}E^{c(y)}H^{(z)}_{1}
 +  
\sum_{x,y}\bar{h}_{(x)(y)}\bar{Q}^{(-x)}\bar{U^c}^{(-y)}\bar{H}^{(-z)}_{2} 
\nn \\
& + 
& \sum_{x,y}\bar{k}_{(x)(y)}\bar{Q}^{(-x)}\bar{D^c}^{(-y)}\bar{H}^{(-z)}_1 
+ \sum_{x,y}\bar{l}_{(x)(y)}\bar{L}^{(-x)}\bar{E^c}^{(-y)}\bar{H}^{(-z)}_1 
\nn \\
& + & \sum_z r_{H^{(z)}_1}\Phi_{H_1}H^{(z)}_1\bar{H}^{(-z)}_1
+ \sum_z r_{H^{(z)}_2}\Phi_{H_2}H^{(z)}_2\bar{H}^{(-z)}_2
\nn \\
& + & \sum_x r_{Q^{(x)}}\Phi_QQ^{(x)}\bar{Q}^{(-x)}
+ \sum_y r_{U^{(y)}}\Phi_{U}U^{c(y)}\bar{U}^{c(-y)}
+ \sum_y r_{D^{(y)}}\Phi_DD^{c(y)}\bar{D}^{c(-y)}
\nn \\
& + & \sum_x r_{L^{(x)}}\Phi_LL^{(x)}\bar{L}^{(-x)}
+ \sum_y r_{E^{(y)}}\Phi_{E}E^{c(y)}\bar{E}^{c(-y)}
\nn \\
& + & \sum_z s_{H^{(z)}_1}\theta_{H_1}H^{(z)}_1\bar{H}^{(-z-1)}_1
+ \sum_z \bar{s}_{H^{(z)}_1}\bar{\theta}_{H_1}{H}^{(z)}_1\bar{H}^{(-z+1)}_1
\nn \\
& + & \sum_z s_{H^{(z)}_2}\theta_{H_2}H^{(z)}_2\bar{H}^{(-z-1)}_2
+ \sum_z \bar{s}_{H^{(z)}_2}\bar{\theta}_{H_2}{H}^{(z)}_2\bar{H}^{(-z+1)}_2
\nn \\
& + & \sum_x s_{Q^{(x)}}\theta_{Q}Q^{(x)}\bar{Q}^{(-x-1)}
+ \sum_x \bar{s}_{Q^{(x)}}\bar{\theta}_{Q}{Q}^{(x)}\bar{Q}^{(-x+1)}
\nn \\
& + & \sum_y s_{U^{(y)}}\theta_{U}U^{c(y)}\bar{U}^{c(-y-1)}
+ \sum_y \bar{s}_{U^{(y)}}\bar{\theta}_{U}{U}^{c(y)}\bar{U}^{c(-y+1)}
\nn \\
& + & \sum_y s_{D^{(y)}}\theta_{D}D^{c(y)}\bar{D}^{c(-y-1)}
+ \sum_y \bar{s}_{D^{(y)}}\bar{\theta}_{D}{D}^{c(y)}\bar{D}^{c(-y+1)}
\nn \\
& + & \sum_x s_{L^{(x)}}\theta_{L}L^{(x)}\bar{L}^{(-x-1)}
+ \sum_x \bar{s}_{L^{(x)}}\bar{\theta}_{L}{L}^{(x)}\bar{L}^{(-x+1)}
\nn \\
& + & \sum_y s_{E^{(y)}}\theta_{E}E^{c(y)}\bar{E}^{c(-y-1)}
+ \sum_y \bar{s}_{E^{(y)}}\bar{\theta}_{E}{E}^{c(y)}\bar{E}^{c(-y+1)}
\label{bigsuperpotential}
\eea
where in the first two lines of Eq.\ref{bigsuperpotential}, $X$ symmetry
requires that $z=-(x+y)$.
Since the fields above are being labeled by their $X$ charges,
the limits of each of the summations will depend on the particular
model under consideration. The family indices range from $i,j=1,\cdots 3$.
However in specific models only a subset of the fields
will be present, and consequently not all of the terms
will be present. For the moment we prefer to keep the values
of $n_Q,n_U,n_D,n_L,n_E$ and $n_{H_1},n_{H_2}$ general, however.
Also, we have not written the most general superpotential allowed under the
gauge symmetry since $\theta_E$ could couple to the vector quarks, for
example. It is possible that $\theta_{Q,U,D,L,E,1,2}$ are identified with just
one superfield and that $\phi_{Q,U,D,L,E,{H_1},H_{2}}$ are also identified
with one superfield (and
similarly for the conjugate singlets).

The one-loop RGEs for the couplings 
$R^h_{ij}$,$R^k_{ij}$,$R^l_{ij}$ are
\bea
\frac{dR^h_{ij}}{dt} & = & 
R^h_{ij}(N_{Q_i} + N_{U^c_j}+N_{H^{(z)}_2}+\tilde{\alpha}b )\nn \\
\frac{dR^k_{ij}}{dt} & = & 
R^k_{ij}(N_{Q_i}  +  N_{D^c_j}+N_{H^{(z)}_1}+\tilde{\alpha}b)\nn \\
\frac{dR^l_{ij}}{dt} & = & 
R^l_{ij}(N_{L_i}  +  N_{E^c_j}+N_{H^{(z)}_1}+\tilde{\alpha}b) \label{genrges}
\eea 
where we have assumed the gauge couplings are approximately equal so that
\beq
\tilde{\alpha} \equiv \frac{5}{3} \tilde{\alpha}_1 = \tilde{\alpha}_2 =
\tilde{\alpha}_3.
\eeq
The full list of wavefunction renormalisations are given
in appendix~2.
The fixed point conditions for the chiral quark and
lepton couplings, 
 $R^h_{ij}$,$R^k_{ij}$,$R^l_{ij}$,
are listed below:
\bea
N_{Q_i} + N_{U^c_j}+N_{H^{(z)}_2}+\tilde{\alpha}b & =& 0\nn \\
N_{Q_i}  +  N_{D^c_j}+N_{H^{(z)}_1}+\tilde{\alpha}b& =&0\nn \\
N_{L_i}  +  N_{E^c_j}+N_{H^{(z)}_1}+\tilde{\alpha}b& =&0
\label{genfixedpointconds}
\eea
We also require a similar fixed point for
the couplings $R^h_{i(y)}$, $R^k_{i(y)}$, $R^l_{i(y)}$,
and $R^h_{(x)j}$, $R^k_{(x)j}$, $R^l_{(x)j}$, that involve
a mixture of chiral and vector fields.
Also we require a fixed point for the couplings
$R^h_{(xy)}$, $R^k_{(xy)}$, $R^l_{(xy)}$, involving
purely vector fields.
Similar fixed point conditions apply to the conjugate vector
couplings, as well as all the singlet couplings.
So every trilinear coupling will have a fixed point condition
which is expressed in terms of the wavefunction renormalisations,
similar to the above conditions. A fixed point is achieved when all 
the conditions are simultaneously satisfied. 
Note this assumes that
none of the $R^{h,k,l}_{ij}$ couplings are zero at the fixed point, another
set of possibilities allowed by Eq.\ref{genrges}. We will not consider this
here since
many of the preceding arguments relied on the dimensionless couplings being
$\sim O(1)$, rather than approximately zero. We merely note that in general
there are $2^n$ fixed points in this multi-dimensional system of $n$
couplings, all but one of
which involve some of the dimensionless couplings being zero.

\subsection{Conditions for Infra-Red Attractiveness}

We now write the RG equations as
\begin{equation}
\frac{d R_i(t)}{d t} = \tilde{\alpha} R_i(t) \left[ (r_i+b) - \sum_j S_{ij}
R_j(t)
\right], \label{RGER}
\end{equation}
where $R_i$ now denotes the ratio of any Yukawa coupling $i$ to the gauge
coupling (squared), as prescribed by Eq.\ref{ratio}.
We have written $r_i\equiv 2 \sum C_{2}(R_x)$, where the sum runs over simple
gauge groups and the representations $R_x$ under those gauge groups, $x$
corresponding to the field that labels $N_x$ of $R_i$ in
Eq.\ref{genrges}. 
The
fixed point condition is then satisfied when the 
right hand side of Eq.\ref{RGER} is zero for all $i$. 
First, we assume that none of the $R_i$ is equal to zero at the fixed point,
in which case
\begin{equation}
\sum_j S_{ij} R_j^* = r_i+b, \label{FPcond}
\end{equation}
where we have denoted the value of $R_j$ at the fixed point as $R_j^*$.
The problem of locating the fixed points becomes a straightforward
problem in linear algebra, albeit involving a large number of dimensions,
corresponding to the large number of trilinear Yukawa couplings.
The fixed points are given in principle by inverting the matrix $S_{ij}$,
\beq
R_i^*=\sum_j (S^{-1})_{ij}(r_j +b) \label{howto}
\eeq
To determine the infra-red stability of the system in Eq.\ref{RGER}, we need
to Taylor expand it around the fixed point given in Eq.\ref{FPcond}.
We can then drop all except the linear terms, the resulting system of which
allows an algebraic solution and can
thus be tested for infra-red stability. 
We therefore make a change of variables to $\rho_i(t) \equiv R_i(t) - R_i^*$.
The RGE Eq.\ref{RGER} then becomes
\beq
\frac{d \rho_i(t)}{d t} = - \tilde{\alpha}(t) (\rho_i(t)+ R_i^*) 
\left[ (r_i+b) - \sum_{j}
S_{ij}
(\rho_j(t)+R_j^*)\right], \label{Taylor}
\eeq
where we have substituted the fixed point values of $R_i^*$ from
Eq.\ref{FPcond}. 
When we drop the quadratic term in Eq.\ref{Taylor} and change the independent
variable from $t$ to $\tilde{\alpha}$ by Eq.\ref{Bfns}, we obtain the
linearised system
\begin{equation}
\frac{d \rho_i(t)}{d \ln \tilde{\alpha}(t)} = \frac{1}{b} R_i^* \sum_j S_{ij}
\rho_j(t). \label{linearised}
\end{equation}
Eq.\ref{linearised} then describes the behaviour of the trajectories as they
approach the fixed point. It has solutions
\begin{equation}
\rho_j(t) = x_j \left( \tilde{\alpha}(t) \right)^{\lambda_j}, \label{solns2}
\end{equation}
where $x_j, \lambda_j$ satisfy the eigenvalue equation
\begin{eqnarray}
\sum_j A_{ij} x_j &=& \lambda_i x_i \nonumber \\
A_{ij} &\equiv& \frac{1}{b} R_i^* S_{ij}. \label{evalue}
\end{eqnarray}
Because the expanded RGE Eq.\ref{linearised} is linear, the general solution
is a linear combination of each $\rho_i$ in Eq.\ref{linearised}.

For $b>0$ as in these models, $\tilde{\alpha}$ decreases with decreasing
renormalisation scale 
$\mu$. For the fixed point to be infra-red stable, we require every eigenvalue
$\lambda_i$ to have a positive real part, since then $\rho_i \rightarrow 0$ as
$\mu$ 
decreases. 
This simply translates to the
condition that
every eigenvalue $\lambda$ of $A$ must possess a positive real
part. Complex eigenvalues always come in complex conjugate pairs, as do their
associated eigenvectors. Writing $\lambda_j \equiv k_j+is_j$, where $k_j,s_j$
are
real, the solution in this case is
\beq
x_j \tilde{\alpha}^{k_j + is_j} + x_j^* \tilde{\alpha}^{k_j-is_j}
= \tilde{\alpha}^{k_j} \left[ x_j \tilde{\alpha}^{i s_j} + x_j^*
\tilde{\alpha}^{-is_j} \right]. \label{spiral} 
\eeq
Eq.\ref{spiral} describes a spiral-like trajectory, the distance to the fixed
point being controlled by $\tilde{\alpha}^{k_j}$. Thus $k_j$ must be positive
for the trajectory to be infra-red stable.

$\lambda_i=0$ corresponds to a direction in coupling space which is neither
attracted nor repelled by the fixed point. 
For each of these directions there
should be one free parameter in the solution to the fixed point equations.
Thus, the dimension of the
null-space of $A$ gives the number of free parameters.
The free parameters embody the information on where a solution lies along this
direction (and are set by the initial boundary conditions). 
In the class of
models presented here, this corresponds to some
information about the string scale being retained at lower energies.
In fact, it can be shown that the null-space of $S$ is the
null-space of $A$ and so the number of zero eigenvalues of $S$ fixes the
number of free parameters.
If the conditions for infra-red stability are not met, the fixed point is
either a saddle point or
an ultra-violet fixed point and so the fixed point will never be achieved at
low energies. We will see in the following specific models,
examples of
infra-red stable and saddle point behaviour. We will also see that the
null-space directions
occur because of degeneracies in the fixed point equations.

\subsection{Example 1: Higgs Mixing Model}

As a first example of the general results,
we calculate the fixed point solutions and the infra-red stability
in a Higgs mixing model similar to that proposed by Ross \cite{GG}. 
In this model there are $n_{H_1}=10,n_{H_2}=10$
copies of the vector representations
\beq
(H^{(z)}_{2}+\bar{H}^{(-z)}_{2}),\ (H^{(z)}_{1}+\bar{H}^{(-z)}_{1}),
\eeq
in the model, which means
$n_2=20$. Ross also included some colour triplets which served the purpose of
increasing the gauge unification scale although not enough to be
consistent alone with string-scale gauge unification. 
We saw in Eq.\ref{equal} that such a model has string scale gauge unification
if $n_S=24$, but we shall ignore
the exotic sexton representations in the following analysis.
The superpotential of the model is then
\bea
W & = &
\sum_{i,j}h_{ij}Q_iU^{c}_jH^{(z)}_{2}
+  \sum_{i,j}k_{ij}Q_iD^c_jH^{(z)}_{1}
 +  \sum_{i,j}l_{ij}L_iE^c_jH^{(z)}_{1}
\nn \\
& + & \sum_{z=-2}^8 r_{H^{(z)}_1}\Phi_{H_1}H^{(z)}_1\bar{H}^{(-z)}_1
+ \sum_{z=-2}^8 r_{H^{(z)}_2}\Phi_{H_2}H^{(z)}_2\bar{H}^{(-z)}_2
\nn \\
& + & \sum_{z=-2}^{7} s_{H^{(z)}_1}\theta_{H_1}H^{(z)}_1\bar{H}^{(-z-1)}_1
+ \sum_{z=-1}^8
\bar{s}_{H^{(z)}_1}\bar{\theta}_{H_1}{H}^{(z)}_1\bar{H}^{(-z+1)}_1
\nn \\
& + & \sum_{z=-2}^7 s_{H^{(z)}_2}\theta_{H_2}H^{(z)}_2\bar{H}^{(-z-1)}_2
+ \sum_{z=-1}^8
\bar{s}_{H^{(z)}_2}\bar{\theta}_{H_2}{H}^{(z)}_2\bar{H}^{(-z+1)}_2.
\label{rossW}
\eea
It is to be understood in the first three terms Eq.\ref{rossW} that
\beq
z=-\mbox{Xcharge}(i^{th}\mbox{~family})-\mbox{Xcharge}(j^{th}\mbox{~family}).
\eeq
Therefore the Higgs which occur in the
interactions with couplings $(h,k,l)_{ij}$ have charges as given below:
\begin{eqnarray}
\left( 
\begin{array}{ccc}
(h,k,l)_{11} H_{1,2}^{(8)} & (h,k,l)_{12}H_{1,2}^{(3)}
&(h,k,l)_{13}H_{1,2}^{(4)}\\    
(h,k,l)_{21} H_{1,2}^{(3)} & (h,k,l)_{22}H_{1,2}^{(-2)}  &
(h,k,l)_{23}H_{1,2}^{(-1)}\\ 
(h,k,l)_{31} H_{1,2}^{(4)} & (h,k,l)_{32}H_{1,2}^{(-1)} & (h,k,l)_{33}H_{1,2}
\end{array}
\right)
\label{Higgsmat}
\end{eqnarray}
The above Higgs having direct couplings are only a subset
of the full list of required Higgses:
\bea
  H_{1,2}^{(8)}, \bar{H}_{1,2}^{(-8)},
 H_{1,2}^{(7)},  \bar{H}_{1,2}^{(-7)},
  H_{1,2}^{(6)},  \bar{H}_{1,2}^{(-6)},
 H_{1,2}^{(5)}, \bar{H}_{1,2}^{(-5)},  
  H_{1,2}^{(4)},  \bar{H}_{1,2}^{(-4)},\nn \\
 H_{1,2}^{(3)},  \bar{H}_{1,2}^{(-3)},
  H_{1,2}^{(2)},  \bar{H}_{1,2}^{(-2)},
 H_{1,2}^{(1)}, \bar{H}_{1,2}^{(-1)},  
  H_{1,2}^{(-1)},  \bar{H}_{1,2}^{(1)},
 H_{1,2}^{(-2)}, \bar{H}_{1,2}^{(2)}   
\label{fullHiggslist}
\eea

The one-loop wavefunction renormalisations from Appendix 2 are:
\bea
N_{Q_i}   & = & \frac{8}{3}\tilde{\alpha_3}
+\frac{3}{2}\tilde{\alpha_2} +\frac{1}{18}\tilde{\alpha_1}
-\sum_j Y^h_{ij} -\sum_j Y^k_{ij}
\nn \\
N_{U^c_j}   & = & \frac{8}{3}\tilde{\alpha_3}
+\frac{8}{9}\tilde{\alpha_1}
-2\sum_i Y^h_{ij}
\nn \\
N_{D^c_j}   & = & \frac{8}{3}\tilde{\alpha_3}
+\frac{2}{9}\tilde{\alpha_1}
-2\sum_i Y^k_{ij}
\nn \\
N_{L_i}   & = &
\frac{3}{2}\tilde{\alpha_2} +\frac{1}{2}\tilde{\alpha_1}
-\sum_j Y^l_{ij}
\nn \\
N_{E^c_j}   & = & 2\tilde{\alpha_1}
-2\sum_i Y^l_{ij}
\eea
The Higgs wavefunction contributions are:
\bea
N_{H^{(z)}_1}   & = &
\frac{3}{2}\tilde{\alpha_2} +\frac{1}{2}\tilde{\alpha_1}
-3\sum_{ij} Y^k_{ij} -\sum_{ij} Y^l_{ij}
\nn \\
& - & Y^{r_{H^{(z)}_1}}-Y^{s_{H^{(z)}_1}}-Y^{\bar{s}_{H^{(z)}_1}}
\nn \\
N_{H^{(z)}_2}   & = &
\frac{3}{2}\tilde{\alpha_2} +\frac{1}{2}\tilde{\alpha_1}
\nn \\
& - & 3\sum_{ij} Y^h_{ij}
-Y^{r_{H^{(z)}_2}}-Y^{s_{H^{(z)}_2}}-Y^{\bar{s}_{H^{(z)}_2}}
\nn \\
N_{\bar{H}^{(-z)}_1}   & = &
\frac{3}{2}\tilde{\alpha_2} +\frac{1}{2}\tilde{\alpha_1}
-Y^{r_{H^{(z)}_1}}-Y^{\bar{s}_{H^{(z+1)}_1}}-Y^{{s}_{H^{(z-1)}_1}}
\nn \\
N_{\bar{H}^{(-z)}_2}   & = &
\frac{3}{2}\tilde{\alpha_2} +\frac{1}{2}\tilde{\alpha_1}
-Y^{r_{H^{(z)}_2}}-Y^{\bar{s}_{H^{(z+1)}_2}}-Y^{{s}_{H^{(z-1)}_2}}
\eea
The wavefunction contributions for the singlets are:
\bea
N_{\Phi_{H_1}}& = & -2\sum_z Y^{r_{H^z_1}},
\
N_{\Phi_{H_2}} = -2\sum_z Y^{r_{H^z_2}}
\nn \\
N_{\theta_{H_1}}& =& -2\sum_z Y^{s_{H^z_1}},
\
N_{\theta_{H_2}} = -2\sum_z Y^{s_{H^z_2}}
\nn \\
N_{\bar{\theta}_{H_1}}& =& -2\sum_z Y^{\bar{s}_{H^z_1}},
\
N_{\bar{\theta}_{H_2}} = -2\sum_z Y^{\bar{s}_{H^z_2}}. \label{wfend}
\eea

Following Ross, we assume a general symmetric form for the matrices $l,k,h$ at
the
string scale and use the fact that the RGEs Eq.\ref{RGER} respect this form.
Like Ross, we also ignore all of the couplings in Eq.\ref{rossW} that do not
involve MSSM states. 
The solutions of the fixed point equations Eq.\ref{howto} applied to this
model are~\cite{GG}
\begin{eqnarray}
R^{h_{ij}} &=& \left(\frac{887}{1728} + \frac{3\ b}{64}\right) \left(
\begin{array}{ccc} 2 & 1 & 1 \\ 1 & 2 & 1 \\ 1 & 1 & 2 \\ \end{array}\right)
\nn \\ 
R^{k_{ij}} &=&  \left(
\begin{array}{ccc}
\frac{1297}{864} + \frac{5\ b}{32} -x -y & y & x\\
y & \frac{1297}{864} + \frac{5\ b}{32} - y - z & z\\
x & z & \frac{1297}{864} + \frac{5\ b}{32} - z - x \\ \end{array}
\right) \nn \\
R^{l_{ij}} &=& \left(\begin{array}{ccc}
-\frac{103}{72} - \frac{b}{8}+3x+3y & \frac{295}{192} + \frac{11\ b}{64} - 3y
& \frac{295}{192} + \frac{11\ b}{64} - 3 x\\
\frac{295}{192} + \frac{11\ b}{64} - 3y & -\frac{103}{72} - \frac{b}{8} +3y+3z
& \frac{295}{192} + \frac{11\ b}{64} - 3z\\
\frac{295}{192} + \frac{11\ b}{64} - 3x & \frac{295}{192} + \frac{11\ b}{64} -
3 z & -\frac{103}{72} - \frac{b}{8} +3x+3z
 \\ \end{array}\right)
\label{rossol}
\end{eqnarray}
Eq.\ref{rossol} shows that there are 3 undetermined parameters at the
fixed point. 
Note that if any of the entries of the matrices in Eq.\ref{rossol} are
negative, the coupling corresponding to the negative solution will tend to
zero. This could violate the assumptions about the fundamental dimensionless
couplings all being of order unity and hence the choice of $X$ charges
required to reproduce the phenomenology. 
We therefore demand that $x,y,z$ must satisfy
the constraints
\bea
\{x+y,x+z,y+x\} &>& \frac{103}{216} + \frac{b}{24} \nn \\
0 < \{x,y,z\} &<& \frac{295}{576}+\frac{11b}{192}. \label{ineq}
\eea
When we check the solution for infra-red stability we
find that the eigenvalues of $A$ are positive for solutions that satisfy the
above conditions.
Thus, the fixed point solution identified is
completely infra-red stable, with three undetermined free parameters.

So far, in the limit that singlets are ignored,
our results are in agreement with those of ref.\cite{GG}.
Although the stability question was not explicitly addressed \cite{GG} we find
that the
fixed point is stable in the infra-red limit so all is well.
Now we must consider the effect of the singlets.
In the Ross model \cite{GG} the result was quoted that
the singlet couplings of the Higgs
Mixing Model are approximately flavour independent. 
We find that this 
is only valid if the extra Higgs doublets which do not have direct
couplings to fermions are ignored. To be explicit,
Ross considered a model with the only extra Higgs states being
\beq
H_{1,2}^{(8)},\bar{H}_{1,2}^{(-8)},H_{1,2}^{(4)},
\bar{H}_{1,2}^{(-4)},H_{1,2}^{(3)},\bar{H}_{1,2}^{(-3)}, 
H_{1,2}^{(2)},\bar{H}_{1,2}^{(-2)},H_{1,2}^{(1)},\bar{H}_{1,2}^{(-1)}.
\eeq
However we saw earlier that the full list of Higgs states in
Eq.\ref{fullHiggslist} is required for correct
Cabbibo mixing and CKM mixing.
The full Higgs mixing model is analysed in Appendix 3 where we
solve the 80 simultaneous equations for 80 unknowns (keeping the
matrices $k,l,h$ symmetric) to determine the predictive
properties of the model. The solution detailed in Appendix 3 shows
that the solution has 27 undetermined parameters. 
We now discard the model
because we have more unconstrained parameters than data points on fermion
masses and mixings. 

\subsection{Example 2: Quark-Line Mixing Model}

We next consider a model 
with mixing along the $Q$ doublet line provided by the extra fields
\beq
Q^{(4)},Q^{(3)},Q^{(2)},Q^{(1)},Q^{(0)},Q^{(-1)},Q^{(-2)},Q^{(-3)},Q^{(-4)}
\eeq
and their partners in the vector-like pair.
Such a model by itself is not expected to be realistic
since it does not account for lepton masses, but it may be regarded
as part of a fuller model such as the $n_Q=9,n_U=4,n_D=10,n_E=8$
example in Tables~\ref{tab:mainresults},\ref{tab:res2}.

The superpotential of this example is explicitly:
\bea
W & = & h_{33}Q_3U^{c}_3H_{2}  
+  h_{(4)1}Q^{(4)}U^{c}_1H_{2}  
+  h_{(-1)2}Q^{(-1)}U^{c}_2H_{2} 
+  h_{(0)3}Q^{(0)}U^{c}_3H_{2}   \nn \\
& + &  \sum_{x=-4}^{4} r_{Q^{(x)}}\Phi_QQ^{(x)}\bar{Q}^{(-x)}
\nn \\
& + &  s_{Q_1}\theta_{Q}Q_1\bar{Q}^{(3)}
+ s_{Q_2}\theta_{Q}Q_2\bar{Q}^{(-2)}
+ s_{Q_3}\theta_{Q}Q_3\bar{Q}^{(-1)}
\nn \\
& +& \bar{s}_{Q_2}\bar{\theta}_{Q}Q_2\bar{Q}^{(0)}
+\bar{s}_{Q_3}\bar{\theta}_{Q}Q_3\bar{Q}^{(1)}
\nn \\
& + & \sum_{x=-4}^{3} s_{Q^{(x)}}\theta_{Q}Q^{(x)}\bar{Q}^{(-x-1)}
+ \sum_{x=-3}^4 \bar{s}_{Q^{(x)}}\bar{\theta}_{Q}{Q}^{(x)}\bar{Q}^{(-x+1)}
\label{modelsuperpotential}
\eea

The wavefunction renormalisations for the chiral quarks and  leptons 
and MSSM Higgs doublets are
(see Appendix 2):
\bea
N_{Q_3}   & = & \frac{8}{3}\tilde{\alpha_3} 
+\frac{3}{2}\tilde{\alpha_2} +\frac{1}{18}\tilde{\alpha_1} 
-Y^h_{33} -Y^{s_{Q_3}}-Y^{\bar{s}_{Q_3}}
\nn \\
N_{Q_2}   & = & \frac{8}{3}\tilde{\alpha_3} 
+\frac{3}{2}\tilde{\alpha_2} +\frac{1}{18}\tilde{\alpha_1} 
-Y^{s_{Q_2}}-Y^{\bar{s}_{Q_2}}
\nn \\
N_{Q_1}   & = & \frac{8}{3}\tilde{\alpha_3} 
+\frac{3}{2}\tilde{\alpha_2} +\frac{1}{18}\tilde{\alpha_1} 
-Y^{s_{Q_1}}
\nn \\
N_{U^c_3}   & = & \frac{8}{3}\tilde{\alpha_3} 
+\frac{8}{9}\tilde{\alpha_1} 
-2Y^h_{33} -2Y^h_{(0)3} 
\nn \\
N_{U^c_2}   & = & \frac{8}{3}\tilde{\alpha_3} 
+\frac{8}{9}\tilde{\alpha_1} 
-2Y^h_{(-1)2} 
\nn \\
N_{U^c_1}   & = & \frac{8}{3}\tilde{\alpha_3} 
+\frac{8}{9}\tilde{\alpha_1} 
-2Y^h_{(4)1} 
\nn \\
N_{H_2}   & = & 
\frac{3}{2}\tilde{\alpha_2} +\frac{1}{2}\tilde{\alpha_1} 
 -  3Y^h_{33} 
-3(Y^h_{(4)1} +Y^h_{(-1)2} +Y^h_{(0)3} )
\nn \\
N_{Q^{(0)}}   & = & \frac{8}{3}\tilde{\alpha_3} 
+\frac{3}{2}\tilde{\alpha_2} +\frac{1}{18}\tilde{\alpha_1} 
- Y^h_{(0)3}
-Y^{r_{Q_{(0)}}}-Y^{s_{Q_{(0)}}}-Y^{\bar{s}_{Q_{(0)}}}
\nn \\
N_{Q^{(-1)}}   & = & \frac{8}{3}\tilde{\alpha_3} 
+\frac{3}{2}\tilde{\alpha_2} +\frac{1}{18}\tilde{\alpha_1} 
- Y^h_{(-1)2}
-Y^{r_{Q_{(-1)}}}-Y^{s_{Q_{(-1)}}}-Y^{\bar{s}_{Q_{(-1)}}}
\nn \\
N_{Q^{(4)}}   & = & \frac{8}{3}\tilde{\alpha_3} 
+\frac{3}{2}\tilde{\alpha_2} +\frac{1}{18}\tilde{\alpha_1} 
- Y^h_{(4)1}
-Y^{r_{Q_{(4)}}}-Y^{\bar{s}_{Q_{(4)}}}
\nn \\
N_{Q^{(1)}}   & = & \frac{8}{3}\tilde{\alpha_3} 
+\frac{3}{2}\tilde{\alpha_2} +\frac{1}{18}\tilde{\alpha_1} 
-Y^{r_{Q_{(1)}}}-Y^{s_{Q_{(1)}}}
\nn \\
N_{Q^{(x\neq 1,0,-1,4)}}   & = & \frac{8}{3}\tilde{\alpha_3} 
+\frac{3}{2}\tilde{\alpha_2} +\frac{1}{18}\tilde{\alpha_1} 
-Y^{r_{Q_{(x)}}}-Y^{s_{Q_{(x)}}}-Y^{\bar{s}_{Q_{(x)}}}
\nn \\
N_{\bar{Q}^{(3)}}   & = & \frac{8}{3}\tilde{\alpha_3} 
+\frac{3}{2}\tilde{\alpha_2} +\frac{1}{18}\tilde{\alpha_1}   
-Y^{r_{Q^{(-3)}}}
-Y^{\bar{s}_{Q^{(-2)}}}-Y^{{s}_{Q^{(-4)}}}-Y^{{s}_{Q_1}}
\nn \\
N_{\bar{Q}^{(-2)}}   & = & \frac{8}{3}\tilde{\alpha_3} 
+\frac{3}{2}\tilde{\alpha_2} +\frac{1}{18}\tilde{\alpha_1}   
-Y^{r_{Q^{(2)}}}
-Y^{\bar{s}_{Q^{(3)}}}-Y^{{s}_{Q^{(1)}}}-Y^{{s}_{Q_2}}
\nn \\
N_{\bar{Q}^{(-1)}}   & = & \frac{8}{3}\tilde{\alpha_3} 
+\frac{3}{2}\tilde{\alpha_2} +\frac{1}{18}\tilde{\alpha_1}   
-Y^{r_{Q^{(1)}}}
-Y^{\bar{s}_{Q^{(2)}}}-Y^{{s}_{Q^{(0)}}}-Y^{{s}_{Q_3}}
\nn \\
N_{\bar{Q}^{(0)}}   & = & \frac{8}{3}\tilde{\alpha_3} 
+\frac{3}{2}\tilde{\alpha_2} +\frac{1}{18}\tilde{\alpha_1}   
-Y^{r_{Q^{(0)}}}
-Y^{\bar{s}_{Q^{(1)}}}-Y^{{s}_{Q^{(-1)}}}-Y^{\bar{s}_{Q_2}}
\nn \\
N_{\bar{Q}^{(1)}}   & = & \frac{8}{3}\tilde{\alpha_3} 
+\frac{3}{2}\tilde{\alpha_2} +\frac{1}{18}\tilde{\alpha_1}   
-Y^{r_{Q^{(-1)}}}
-Y^{\bar{s}_{Q^{(0)}}}-Y^{{s}_{Q^{(-2)}}}-Y^{\bar{s}_{Q_3}}
\nn \\
N_{\bar{Q}^{(x\neq 3,-2,-1,0,1)}}   & = & \frac{8}{3}\tilde{\alpha_3} 
+\frac{3}{2}\tilde{\alpha_2} +\frac{1}{18}\tilde{\alpha_1}   
-Y^{r_{Q^{(-x)}}}
-Y^{\bar{s}_{Q^{(-x+1)}}}-Y^{{s}_{Q^{(-x-1)}}}
\nn \\
N_{\Phi_{Q}} & = & -6\sum_{x=-4}^4 Y^{r_{Q^{(x)}}},
\nn \\
N_{\theta_Q} &= & -6\sum_{x=-4}^3 Y^{s_{Q^(x)}}
-6(Y^{s_{Q_1}}+Y^{s_{Q_2}}+Y^{s_{Q_3}}),
\nn \\
N_{\bar{\theta}_{Q}} & = & -6\sum_{x=-3}^4 Y^{\bar{s}_{Q^{(x)}}}
-6( Y^{\bar{s}_{Q_2}}+Y^{\bar{s}_{Q_3}}).
\label{wavefn}
\eea

The fixed point conditions for the
couplings $R^h_{33}$,$R^h_{xj}$, 
$R^{s_{Q_i}}$,$R^{\bar{s}_{Q_i}}$,
$R^{r_{Q^{(x)}}}$, 
$R^{s_{Q^{(x)}}}$, 
$R^{\bar{s}_{Q^{(x)}}}$, are:
\bea
N_{Q_3} + N_{U^c_3}+N_{H_2}+\tilde{\alpha}b & =& 0 \label{rh33}\\
N_{Q^{(4)}}  +  N_{U^c_1}+N_{H_2}+\tilde{\alpha}b& =&0 \label{rhxj}\\
N_{Q^{(-1)}}  +  N_{U^c_2}+N_{H_2}+\tilde{\alpha}b& =&0 \\
N_{Q^{(0)}}  +  N_{U^c_3}+N_{H_2}+\tilde{\alpha}b& =&0 \\
N_{{\theta}_{Q}}+N_{Q_1}+N_{\bar{Q}^{(3)}}
+\tilde{\alpha}b & =& 0  \\
N_{{\theta}_{Q}}+N_{Q_2}+N_{\bar{Q}^{(-2)}}
+\tilde{\alpha}b & =& 0  \\
N_{{\theta}_{Q}}+N_{Q_3}+N_{\bar{Q}^{(-1)}}
+\tilde{\alpha}b & =& 0  \\
N_{{\bar{\theta}}_{Q}}+N_{Q_2}+N_{\bar{Q}^{(0)}}
+\tilde{\alpha}b & =& 0  \\
N_{{\bar{\theta}}_{Q}}+N_{Q_3}+N_{\bar{Q}^{(1)}}
+\tilde{\alpha}b & =& 0  \\
N_{\Phi_{Q}}+N_{Q^{(x)}}+N_{\bar{Q}^{(-x)}}
+\tilde{\alpha}b & =& 0, \ x \in \{-4,\ldots,4\}\\
N_{\theta_Q}+N_{Q^{(x)}}+N_{\bar{Q}^{(-x-1)}}
+\tilde{\alpha}b & =& 0, \ x \in \{3,\ldots,-4\} \\
N_{\bar{\theta}_Q}+N_{Q^{(x)}}+N_{\bar{Q}^{(-x+1)}}
+\tilde{\alpha}b & =& 0, \ x \in \{4,\ldots,-3\} \label{last}
\label{modelfixedpointconds}
\eea

If we were to ignore the contribution of the singlet
sector then the fixed point equations for the couplings
$R^h_{33}$,$R^h_{xj}$, Eqs.\ref{rh33} and~\ref{rhxj},
lead to the matrix equation:
\beq
\sum_j S_{ij} Y_j = 
\left(
\begin{array}{cccc}
6 & 5  & 3 & 3 \\
5 & 6  & 3 & 3 \\
3 & 3  & 6 & 3 \\
3 & 3  & 3 & 6
\end{array}
\right)
\left(
\begin{array}{c}
Y^h_{33}\\
Y^h_{(0)3}\\
Y^h_{(-1)2}\\
Y^h_{(4)1}\\
\end{array}
\right)
=
\left(
\begin{array}{c}
1 \\
1 \\
1 \\
1 \\
\end{array}
\right)
(r^{QUH_2}+b)\tilde{\alpha}
\label{modelmatrixeqn}
\eeq
where $r^{QUH_2}=88/9$, and we have assumed 
all the gauge couplings to be equal.
Upon inverting the matrix we find,
\begin{eqnarray}
\left(
\begin{array}{c}
Y^h_{33}\\
Y^h_{(0)3}\\
Y^h_{(-1)2}\\
Y^h_{(4)1}\\
\end{array}
\right)
=
\left(
\begin{array}{c}
3 \\
3 \\
5 \\
5 \\
\end{array}
\right)
\frac{(r^{QUH_2}+b)}{63}\tilde{\alpha}
\label{modelmatrixsln}
\end{eqnarray}
for the fixed point solutions of the couplings.
Note that this solution has
a global $SU(2)$ flavour
symmetry in the Yukawa couplings of the two lightest
families, unlike the Higgs mixing model
for example \cite{GG}. The reason that it is present in this model
is that there is a single Higgs doublet which is common to all
the fixed point equations, as compared to the Higgs mixing model
where a different Higgs couples in each entry of the Yukawa matrix.
When the singlets are included
they will explicitly break the global $SU(2)$ flavour symmetry,
as we discuss below. 
Note that in some recent models, such a symmetry is assumed
as a starting point \cite{BH}. We then checked that the system of RGEs in
Eq.\ref{genfixedpointconds} is infra-red stable in this case by determining
the eigenvalues of the matrix $A_{ij}$.
These come out to be 
$b+r^{QUH_2}$, $(b+r^{QUH_2})/21$, $5(b+r^{QUH_2})/21$, $5(b+r^{QUH_2})/21$,
so 
the fixed point is encouragingly infra-red stable in all four independent
directions.

Once the singlets are included the above fixed point
in Eq.\ref{modelmatrixsln} are be modified.
If we return to Eqs.\ref{rh33}-\ref{last}
we see that there are the same number of equations as
unknowns, so the whole system may be regarded as a large
matrix which may be inverted along the above lines.
From Eqs.\ref{rh33}-\ref{last} the following
relations may be obtained, 
\bea
N_{Q_3} & = & N_{Q^{(0)}}, \ N_{Q_2}=N_{Q^{(1)}}, \ N_{Q_1}=N_{Q^{(-4)}},  
\\
N_{\Phi_{Q}} & = & \frac{1}{2}(N_{\theta_Q}+N_{\bar{\theta}_Q} )
\\
N_{\theta_Q}-N_{\Phi_{Q}} & = &
N_{Q^{(x+1)}}  -  N_{Q^{(x)}}=
N_{\bar{Q}^{(-x+1)}}-N_{\bar{Q}^{(-x)}}
\nn \\
& = & N_{Q_2}-N_{Q_3}=
N_{U^c_2}-N_{U^c_3}
\nn \\
 & = & \frac{1}{4}(N_{Q_3}-N_{Q_1})=
\frac{1}{4}(N_{U^c_3}-N_{U^c_1})
\nn \\
 & = & \frac{1}{5}(N_{Q_2}-N_{Q_1})=
\frac{1}{5}(N_{U^c_2}-N_{U^c_1}) 
\label{third}
\eea
These relations are formally quite model-independent: they apply to any model
with $n_Q=9$ provided Eq.\ref{chargecond} holds, regardless of the number of
additional states.
However the implications of these relations will depend on the particular
model under consideration since the wavefunction renormalisations 
have model dependence. For instance in this particular example,
we can immediately see that the previously obtained fixed point
based on ignoring the effect of the singlets is not 
consistent with these equalities.
For example it would imply $(N_{U^c_2}-N_{U^c_1})\propto (5-5)=0$
and $(N_{U^c_3}-N_{U^c_1})\propto (9-10)\neq 0$, although the two relations
are approximately consistent. 

In Appendix~4 we give the fixed point of this example,
including the singlets.
The fixed point solution has 9 undetermined parameters and so
again, we see the predictivity of the model is severely lowered by the
inclusion of the singlet couplings.
This behaviour is rather similar to that encountered in the Higgs 
mixing model, and we therefore expect that it may be a general 
feature of models of this kind, once the singlets are included. 
There is also another serious problem with the fixed point solution, given
that it is
impossible to pick values of the 9 free parameters that predict all of the
constrained couplings to be positive. Thus, the fixed point identified cannot
be realised by nature as one or more couplings will be forced toward zero in
any infra-red stable solution.

\section{Conclusions}

We have explored a scenario in which 
the minimal supersymmetric standard model (MSSM)
is valid up to an energy scale of 
$\sim 10^{16}$ GeV, but that above this scale the theory is 
supplemented by extra 
vector-like representations of the gauge group,
plus a gauged $U(1)_X$ family symmetry.
The basic idea of our approach is that the extra heavy matter
above the scale $\sim 10^{16}$ GeV may be used in two different
ways: (1) to allow (two-loop) gauge 
coupling unification at the string scale; (2) to mix with
quarks, leptons and Higgs fields via spaghetti diagrams
and so lead to phenomenologically
acceptable Yukawa textures.

We considered models in which there are enough heavy vector representations to
give every effective MSSM-type Yukawa coupling a non-zero value.
Using this constraint (detailed by Eqs.\ref{up}-\ref{lepton}), plus the
further condition that
the mass scale must not be too far below the string scale,
we performed a two-loop string gauge unification analysis which
yielded 8 models that satisfy these conditions for $\tan \beta=43$ and 2 for
$\tan \beta=5$ in Tables~\ref{tab:mainresults} and~\ref{tab:res2}. 
For example $n_Q=9,n_U=4,n_D=10,n_E=8$ (all other $n_i$ zero) satisfies the
constraints of string unification independently of $\tan \beta$,
and also has enough heavy matter to enable Yukawa textures to be
generated via spaghetti diagrams. An example of a different solution 
is $n_2=20,n_S=24$ where $n_2=20$ may be interpreted as being due to
$n_{H_1}=10,n_{H_2}=10$ as required in the Higgs mixing model.

Because the dimensionless couplings are of order 1 and because the RG
running of the gauge couplings above $M_I$ is steep, one might hope the
dimensionless couplings would be forced toward numerical values approximating a
fixed point at $M_I$. This would allow us to make numerical estimates of the
values of the fermion masses and mixings at low energy.
Motivated by these considerations 
we constructed the superpotential for a general model
involving intermediate matter, and various Standard Model singlets that
provide the U(1)$_X$ breaking. We obtained 
the one-loop RGEs for the general model, and then obtained 
conditions for stability of the fixed point, showing that
the direction of stability in terms of the renormalisation scale depended on
the eigenvalues of the coefficient matrix of the fixed point equations.

Having discussed the general case, 
we then investigated the fixed point of two examples in some detail:
a Higgs-line mixing model with $n_{H_1}=10,n_{H_2}=10$
and a quark-line mixing model with $n_Q=9$.
Both models have infra-red stable fixed points in the approximation
that the couplings involving the singlets are ignored.
However when the singlets are included in the analysis
we found that the number of undetermined parameters grows
(from 3 to 27 in the Higgs-line mixing model, and from
0 to 9 in the quark-line mixing model).
The spirit of the fixed point approach is that the model's couplings will
focus down on a particular set of values at a lower energy scale independent
of the starting
conditions at a higher energy scale.
In the full Higgs-line mixing model, the predictive fixed point solution is
only 
valid given certain constraints on the values of the couplings at high energy,
and so this
seems to be in opposition to the spirit of the fixed point approach.
In the full quark-line mixing model, the fixed point solution demands a
negative value for at least one of the dimensionless coupling (squared)
$R_i$.
Any $R_i$ that have negative fixed point solutions will actually tend toward
zero,
violating the initial assumption of order one dimensionless couplings that
led to the construction of the $U(1)_X$ model. If the full Higgs-mixing model
is to have a solution for which the couplings squared are all positive, a 
complicated set of constraints must hold upon the boundary conditions at the
higher energy scale.
While the examples investigated here predict
rough orders of magnitude for the fermion masses and mixings, it appears that
detailed attempts to make them predictive by the analysis of infra-red fixed
points is problematic as the above cases illustrate. It is unclear how to
avoid these
pitfalls in the construction of predictive models without the inclusion of
some extra symmetry to increase predictivity, and/or boundary conditions
provided by an explicit string model.

The lesson learned is clear: it is not in general
appropriate to ignore the singlet couplings which must be
incorporated fully into the analysis.
The most predictive scenario would be one in which the Yukawa
couplings depend only upon $\tilde{\alpha}$ in the fixed point solution. 
It is not yet clear how one could pick a model in which this is likely to be
true, or how one could pick a model that possesses a completely infra-red
stable fixed point with all Yukawa couplings non-zero. 
However it is possible that such a model is contained in the
subset of the models in
Table~\ref{tab:mainresults} which have not all been analysed
in detail because of their individual algebraic complexity.
The search for a completely realistic model, and the detailed comparison
of its low energy predictions to data, is beyond the scope of this paper.

The idea of being able to predict the entire fermion mass spectrum
in terms of one or two free parameters is an exciting prospect,
and we hope that the general results of the present paper will be
helpful in this endeavour. 
While we were not able to make the models
quantitatively predictive in the fermion mass and mixing sector,
they simultaneously
provide an explanation for string gauge unification and the hierarchies in the
fermion mass sector.

\vspace{\baselineskip}
\noindent{\large \bf Acknowledgments}
\vspace{\baselineskip}

B.A. would like to thank M. Seymour and R. Thorne for discussions on the
four-loop QCD beta function. The work of S.F.K. is partially
supported by PPARC grant number GR/K55738.

\newpage
\noindent{\large \bf
Appendix 1: Two-Loop Renormalisation of MSSM Plus Intermediate States}
\vspace{\baselineskip}

We derived the following RGEs for the two-loop evolution of the gauge
couplings and third family
Yukawa couplings in the case of additional
intermediate matter from ref.~\cite{MandV}. 
Note that we have neglected other Yukawa
couplings as an approximation. 
This
a good approximation for the bulk of the running which is between $M_I$ and 1
TeV, where the intermediate states have been integrated out of the effective
field theory and the effective Yukawa couplings are all much less than 1 apart
for $\lambda_{t,b,\tau}$.
As we are not considering neutrino masses in this analysis, we assume that
there are no neutrino Yukawa couplings in the effective field theory being
considered. 
For now we must assume there are no extra
couplings between the superfields of the MSSM and the extra matter for reasons
of simplicity and computing time.
Naively one might
expect these couplings to only change the results slightly because they
decouple after
less than  2 orders of magnitude in renormalisation running and because they
only 
affect the running of the gauge couplings at the two-loop level.
Nonetheless, it should be borne in mind that these couplings could 
influence
the results, particularly in view of the fact that these dimensionless
couplings are expected to be of order 1. 
The equations are valid in the
$\o.{DR}$ scheme. 
\bea
16 \pi^2 \frac{d g_1}{d t'} &=&
g_1^3 \left[ \frac{33}{5} + 
\frac{n_Q}{5} + \frac{8 n_U}{5} + \frac{2n_D}{5}
+ \frac{n_S}{10} + \frac{3}{5} n_2 + \frac{6}{5} n_E 
+ \r. \nn \\ && \l.
\frac{1}{16 \pi^2} \left( - \frac{26}{5} \lambda_t^2 - \frac{14}{5}
\lambda_b^2 - \frac{18}{5} \lambda_\tau^2 +
\frac{36}{25} g_1^2 \left( 
\frac{199}{25} + 
\frac{n_Q}{108} + \frac{n_2}{4}  +
\r. \r. \r.\nn \\ && \l. 
\frac{2n_D + 32
n_U}{27} +
\frac{2n_S}{43} + 2n_E \right)+ g_2^2 \left( \frac{27}{5} + \frac{3 n_Q}{5} +
\frac{9  n_2}{5} \right) + 
\nn \\ && \l. \l. \  
g_3^2 \frac{16}{5} \left( \frac{88}{5} + \frac{n_Q}{3} +
\frac{n_S}{6} +
\frac{8 n_U + 2n_D}{3} \right) \right)
\right]
\nn \\
16 \pi^2 \frac{d g_2}{d t'} &=&
g_2^3 \left[ 1+3n_Q  +n_2+
\frac{1}{16 \pi^2}  \left( 
- 6\lambda_t^2 - 6\lambda_b^2 - 2\lambda_\tau^2 
+g_2^2 \left( 25 + 
 \r. \r. \r. \nn \\
&& \l. \l. \l.
7 (3 n_Q + n_2)
\right) +
\frac{3}{10} g_1^2 \left( 6 + \frac{2n_Q}{3} + 2n_2 \right)
+ 8 g_3^2 \left(2n_Q + 3\right) \right) \right]
\nn \\
16 \pi^2 \frac{d g_3}{d t'} &=&
g_3^3 \left[ -3 +  2n_Q + n_U + n_D + n_S+
\frac{1}{16 \pi^2} \left( 
- 4\lambda_t^2 - 4\lambda_b^2 +
g_3^2 \left(14 +  \r. \r. \r. \nn \\ && 
\l. 
\frac{68}{3} n_Q + \frac{34}{3} n_S+
\frac{34}{3} n_U + \frac{34}{3} n_D \right) +  
g_1^2 \left( \frac{11}{5} +
\frac{2}{15} n_Q + \frac{16}{15} n_U + \r.
\nn \\ &&  \l. \l. \l.
\frac{4}{15}n_D + 
\frac{2}{30}n_S \right)  
+ g_2^2 \left( 9 + 6 n_Q \right) \right) \right]
\nn \\
16 \pi^2 \frac{d \lambda_t}{d t'} &=&
\lambda_t \left[ 6 \lambda_t^2 + \lambda_b^2 - \frac{13}{15} g_1^2 - 3 g_2^2 -
\frac{16}{3} g_3^2 +
\frac{1}{16 \pi^2} 
\left(g_1^4 \frac{2743}{450}+
 \r. \r. \nn \\
&&\l.  \l. \
g_2^4 \frac{15}{2} - g_3^4 \frac{16}{9} +
\frac{136}{45} g_1^2 g_3^2 + g_1^2 g_2^2 +
8 g_2^2 g_3^2 + \lambda_t^2 \left(
\frac{6}{5} g_1^2 +
 \r. \r. \r. \nn \\
&& \l. \l. \l.
 6 g_2^2 + 16 g_3^2 \right) + \frac{2}{5} \lambda_b^2 g_1^2
- 22 \lambda_t^4 -
5 \lambda_b^2 \lambda_t^2 - 5 \lambda_b^4 - \lambda_b^2
\lambda_\tau^2 \right) \right]
\nn \\
16 \pi^2 \frac{d \lambda_b}{d t'} &=&
\lambda_b \left[ 6 \lambda_b^2 + \lambda_t^2 + \lambda_\tau^2 - \frac{7}{15}
g_1^2 - 3 g_2^2 - \frac{16}{3} g_3^2 + \frac{1}{16 \pi^2} \left(
\frac{287}{90}
g_1^4 + 
\r. \r. \nn \\ &&
\frac{15}{2} g_2^4 - \frac{17}{9} g_3^4 + g_1^2 g_2^2 + \frac{8}{9}
g_1^2 g_3^2 + 8 g_2^2 g_3^2 + \frac{4}{5} \lambda_t^2 g_1^2 + \lambda_b^2
\left( \frac{2}{5} g_1^2  +
\r. \nn \\&&
\l. \l. \l.	
 6 g_2^2 + 16 g_3^2 \right) + \frac{6}{5}
\lambda_\tau^2 g_1^2 - 22 \lambda_b^2 - 5 \lambda_t^2 \lambda_b^2 - 3
\lambda_b^2 \lambda_\tau^2 - 3 \lambda_\tau^4 - 5 \lambda_t^4 \right) \right]
\nn \\
16 \pi^2 \frac{d \lambda_\tau}{d t'} &=&
\lambda_\tau \left[ 4 \lambda_\tau^2 + 3 \lambda_b^2 - \frac{9}{5} g_1^2 - 3
g_2^2 +
\frac{1}{16 \pi^2} \left( \frac{27}{2} g_1^4 + \frac{15}{2} g_2^4 +
\frac{9}{2} g_1^2 g_2^2 
\r. \r. \nn \\&& 
+ \lambda_b^2 \left( 16 g_3^2 - \frac{2}{5} g_1^2
\right) + \lambda_\tau^2 \left( \frac{6}{5} g_1^2 + 6 g_2^2 \right) - 3
\lambda_b^2 \lambda_t^2 - 9 \lambda_b^4 
\nn \\&& \l. \l.
- 9 \lambda_b^2 \lambda_\tau^2 - 10
\lambda_\tau^4 \right) \right] \label{RGEs}
\eea
where $t'=\ln \mu$ and $\mu$ is the $\o.{DR}$ renormalisation scale. 

We now detail the matching conditions between the $\o.{DR}$ and the
$\o.{MS}$ schemes~\cite{MVmatch}:
\bea
g_2^{\o.{DR}} (m_t) &=& \frac{g_2^{\o.{MS}} (m_t)}{1 - {g_2^{\o.{MS}}(m_t)}^2
/ 48 \pi^2} \nn \\
g_3^{\o.{DR}} (m_t) &=& \frac{g_3^{\o.{MS}} (m_t)}{1 - {g_3^{\o.{MS}}(m_t)}^2
/ 32 \pi^2} \nn \\
\lambda_{t,b}^{\o.{DR}} (m_t) &=& \frac{\lambda_{t,b}^{\o.{MS}} (m_t)}{1 +
{g_3^{\o.{MS}}(m_t)}^2 / 12 \pi^2 - {3g_2^{\o.{MS}}(m_t)}^2 / 128 \pi^2},
\label{DRMSmatch}
\eea
where the superscripts denote what scheme the quantity is evaluated in. To a
good approximation, $g_1^{\o.{DR}}=g_1^{\o.{MS}}$ and
$\lambda_\tau^{\o.{DR}}=\lambda_\tau^{\o.{MS}}$.
Our running value of $m_t$ is determined by
\beq
m_t^{phys} = m_t(m_t) (1+\frac{g_3^2(m_t)}{3\pi^2})
\eeq
where $m_t^{phys}$ is the value extracted from experiment. For
$m_t^{phys}=180$ GeV~\cite{pdb} and central values of $\alpha_S(M_Z)$, we
obtain
$m_t(m_t)=166$ GeV.

\vspace{\baselineskip}
\noindent{\large \bf Appendix 2: 
One-loop wavefunction renormalisation of the general
model}
\vspace{\baselineskip}

Here we give the wavefunction renormalisation contributions to the
RGEs for the general model 
in Eq.\ref{bigsuperpotential}.
In the following equations for the wavefunction renormalisation of $N_i$, all
sums are intended to be over couplings that multiply the field $i$ in
Eq.\ref{bigsuperpotential}:
\bea
N_{Q_i}   & = & \frac{8}{3}\tilde{\alpha_3} 
+\frac{3}{2}\tilde{\alpha_2} +\frac{1}{18}\tilde{\alpha_1} 
-\sum_j Y^h_{ij} -\sum_j Y^k_{ij} 
-\sum_y Y^h_{iy} -\sum_y Y^k_{iy} 
-Y^{s_{Q_i}}-Y^{\bar{s}_{Q_i}}
\nn \\
N_{U^c_j}   & = & \frac{8}{3}\tilde{\alpha_3} 
+\frac{8}{9}\tilde{\alpha_1} 
-2\sum_i Y^h_{ij} 
-2\sum_x Y^h_{xj} 
-Y^{s_{U_j}}-Y^{\bar{s}_{U_j}}
\nn \\
N_{D^c_j}   & = & \frac{8}{3}\tilde{\alpha_3} 
+\frac{2}{9}\tilde{\alpha_1} 
-2\sum_i Y^k_{ij} 
-2\sum_x Y^k_{xj} 
-Y^{s_{D^c_j}}-Y^{\bar{s}_{D^c_j}}
\nn \\
N_{L_i}   & = & 
\frac{3}{2}\tilde{\alpha_2} +\frac{1}{2}\tilde{\alpha_1} 
-\sum_j Y^l_{ij} 
-\sum_y Y^l_{iy} 
-Y^{s_{L_i}}-Y^{\bar{s}_{L_i}}
\nn \\
N_{E^c_j}   & = & 2\tilde{\alpha_1} 
-2\sum_i Y^l_{ij} 
-2\sum_x Y^l_{xj} 
-Y^{s_{E^c_j}}-Y^{\bar{s}_{E^c_j}} \label{wf1start}
\eea

The wavefunction renormalisations for the vector states are:
\bea
N_{Q^{(x)}}   & = & \frac{8}{3}\tilde{\alpha_3} 
+\frac{3}{2}\tilde{\alpha_2} +\frac{1}{18}\tilde{\alpha_1} 
\nn \\
& -& \sum_j Y^h_{xj} -\sum_j Y^k_{xj} 
-\sum_y Y^h_{xy} -\sum_y Y^k_{xy} 
-Y^{r_{Q_x}}-Y^{s_{Q_{(x)}}}-Y^{\bar{s}_{Q_{(x)}}}
\nn \\
N_{U^{c(y)}}   & = & \frac{8}{3}\tilde{\alpha_3} 
+\frac{8}{9}\tilde{\alpha_1} 
-2\sum_i Y^h_{iy} 
-2\sum_x Y^h_{xy} 
-Y^{r_{U^{c(y)}}}-Y^{s_{U^{c(y)}}}-Y^{\bar{s}_{U^{c(y)}}}
\nn \\
N_{D^{c(y)}}   & = & \frac{8}{3}\tilde{\alpha_3} 
+\frac{2}{9}\tilde{\alpha_1} 
-2\sum_i Y^k_{iy} 
-2\sum_x Y^k_{xy} 
-Y^{r_{D^{c(y)}}}-Y^{s_{D^{c(y)}}}-Y^{\bar{s}_{D^{c(y)}}}
\nn \\
N_{L^{(x)}}   & = & 
\frac{3}{2}\tilde{\alpha_2} +\frac{1}{2}\tilde{\alpha_1} 
-\sum_j Y^l_{xj} 
-\sum_y Y^l_{xy} 
-Y^{r_{L_x}}-Y^{s_{L_{(x)}}}-Y^{\bar{s}_{L_{(x)}}}
\nn \\
N_{E^{c(y)}}   & = & 2\tilde{\alpha_1} 
-2\sum_i Y^l_{iy} 
-2\sum_x Y^l_{xy} 
-Y^{r_{E^{c(y)}}}-Y^{s_{E^{c(y)}}}-Y^{\bar{s}_{E^{c(y)}}}
\nn \\
N_{H^{(z)}_1}   & = & 
\frac{3}{2}\tilde{\alpha_2} +\frac{1}{2}\tilde{\alpha_1}
-3\sum_{ij} Y^k_{ij} -\sum_{ij} Y^l_{ij} 
-3\sum_{iy} Y^k_{iy} -\sum_{iy} Y^l_{iy}
\nn \\
& - & 3\sum_{xj} Y^k_{xj} -\sum_{xj} Y^l_{xj} 
-3\sum_{xy} Y^k_{xy} -\sum_{xy} Y^l_{xy} 
-Y^{r_{H^{(z)}_1}}-Y^{s_{H^{(z)}_1}}-Y^{\bar{s}_{H^{(z)}_1}} 
\nn \\
N_{H^{(z)}_2}   & = & 
\frac{3}{2}\tilde{\alpha_2} +\frac{1}{2}\tilde{\alpha_1} 
\nn \\
& - & 3\sum_{ij} Y^h_{ij} 
-3\sum_{iy} Y^h_{iy} 
-3\sum_{xj} Y^h_{xj} 
-3\sum_{xy} Y^h_{xy} 
-Y^{r_{H^{(z)}_2}}-Y^{s_{H^{(z)}_2}}-Y^{\bar{s}_{H^{(z)}_2}}
\nn \\
N_{\bar{Q}^{(-x)}}   & = & \frac{8}{3}\tilde{\alpha_3} 
+\frac{3}{2}\tilde{\alpha_2} +\frac{1}{18}\tilde{\alpha_1} 
\nn \\
& - &  \sum_y Y^{\bar{h}}_{xy} -\sum_y Y^{\bar{k}}_{xy} 
-Y^{r_{Q_x}}
-Y^{\bar{s}_{Q_{(x+1)}}}-Y^{{s}_{Q_{(x-1)}}}-Y^{{s}_{Q_i}}-Y^{\bar{s}_{Q_i}}
\nn \\
N_{\bar{U}^{c(-y)}}   & = & \frac{8}{3}\tilde{\alpha_3} 
+\frac{8}{9}\tilde{\alpha_1} 
-2\sum_x Y^{\bar{h}}_{xy} 
-Y^{r_{U^{c(y)}}}-Y^{\bar{s}_{U^{c(y+1)}}}-Y^{{s}_{U^{c(y-1)}}}
-Y^{{s}_{U_j}}-Y^{\bar{s}_{U_j}}
\nn \\
N_{\bar{D}^{c(-y)}}   & = & \frac{8}{3}\tilde{\alpha_3} 
+\frac{2}{9}\tilde{\alpha_1} 
-2\sum_x Y^{\bar{k}}_{xy} 
-Y^{r_{D^{c(y)}}}-Y^{\bar{s}_{D^{c(y+1)}}}-Y^{{s}_{D^{c(y-1)}}}
-Y^{{s}_{D_j}}-Y^{\bar{s}_{D_j}}
\nn \\
N_{\bar{L}^{(-x)}}   & = & 
\frac{3}{2}\tilde{\alpha_2} +\frac{1}{2}\tilde{\alpha_1} 
-\sum_y Y^{\bar{l}}_{xy} 
-Y^{r_{L_x}}-Y^{\bar{s}_{L_{x+1}}}-Y^{{s}_{L_{(x-1)}}}
-Y^{{s}_{L_j}}-Y^{\bar{s}_{L_j}}
\nn \\
N_{\bar{E}^{c(-y)}}   & = & 2\tilde{\alpha_1} 
-2\sum_x Y^{\bar{l}}_{xy} 
-Y^{r_{E^{c(y)}}}-Y^{\bar{s}_{E^{c(y+1)}}}-Y^{{s}_{E^{c(y-1)}}}
-Y^{{s}_{E_j}}-Y^{\bar{s}_{E_j}}
\nn \\
N_{\bar{H}^{(-z)}_1}   & = & 
\frac{3}{2}\tilde{\alpha_2} +\frac{1}{2}\tilde{\alpha_1} 
-3\sum_{xy} Y^{\bar{k}}_{xy} -\sum_{xy} Y^{\bar{l}}_{xy} 
-Y^{r_{H^{(z)}_1}}-Y^{\bar{s}_{H^{(z+1)}_1}}-Y^{{s}_{H^{(z-1)}_1}}
\nn \\
N_{\bar{H}^{(z)}_2}   & = & 
\frac{3}{2}\tilde{\alpha_2} +\frac{1}{2}\tilde{\alpha_1} 
-3\sum_{xy} Y^{\bar{h}}_{xy} 
-Y^{r_{H^{(z)}_2}}-Y^{\bar{s}_{H^{(z+1)}_2}}-Y^{{s}_{H^{(z-1)}_2}}
\eea
The wavefunction contributions for the singlets are:
\bea
N_{\Phi_{Q}} & = & -6\sum_x Y^{r_{Q^x}},\
N_{\Phi_{U^c}} =   -3\sum_y Y^{r_{U^{cy}}},\ 
N_{\Phi_{D^c}}=-3\sum_y Y^{r_{D^{cy}}}
\nn \\
N_{\Phi_{L}} &  = & -2\sum_x Y^{r_{L^x}},
\
N_{\Phi_{E^c}} = -\sum_y Y^{r_{E^{cy}}},
\nn \\
N_{\Phi_{H_1}}& = & -2\sum_z Y^{r_{H^z_1}},
\
N_{\Phi_{H_2}} = -2\sum_z Y^{r_{H^z_2}}
\nn \\
N_{\theta_Q} &= & -6\sum_x Y^{s_{Q^x}}-6\sum_i Y^{s_{Q_i}},
\
N_{\theta_{U^c}} = -3\sum_y Y^{s_{U^{cy}}}-3\sum_j Y^{s_{U_j}},
\nn \\
N_{\theta_{D^c}}& = & -3\sum_y Y^{s_{D^{cy}}}-3\sum_j Y^{s_{D_j}},
\nn \\
N_{\theta_L} & = & -2\sum_x Y^{s_{L^x}}-2\sum_i Y^{s_{L_i}},
\
N_{\theta_{E^c}} = -\sum_y Y^{s_{E^{cy}}}-\sum_j Y^{s_{E_j}},
\nn \\
N_{\theta_{H_1}}& =& -2\sum_z Y^{s_{H^z_1}},
\
N_{\theta_{H_2}} = -2\sum_z Y^{s_{H^z_2}}
\nn \\
N_{\bar{\theta}_Q} &= & -6\sum_x Y^{\bar{s}_{Q^x}}-6\sum_i Y^{\bar{s}_{Q_i}},
\
N_{\bar{\theta}_{U^c}} = -3\sum_y Y^{\bar{s}_{U^{cy}}}
-3\sum_j Y^{\bar{s}_{U_j}},
\nn \\
N_{\bar{\theta}_{D^c}}& = & -3\sum_y Y^{\bar{s}_{D^{cy}}}
-3\sum_j Y^{\bar{s}_{D_j}},
\nn \\
N_{\bar{\theta}_L} & = & -2\sum_x Y^{\bar{s}_{L^x}}-2\sum_i Y^{\bar{s}_{L_i}},
\
N_{\bar{\theta}_{{E}^c}} = -\sum_y Y^{\bar{s}_{E^{cy}}}-\sum_j Y^{\bar{s}_{E_j}},
\nn \\
N_{\bar{\theta}_{H_1}}& =& -2\sum_z Y^{\bar{s}_{H^z_1}},
\
N_{\bar{\theta}_{H_2}} = -2\sum_z Y^{\bar{s}_{H^z_2}} \label{wf1end}
\eea

\vspace{\baselineskip}
\noindent{\large \bf Appendix 3: 
Fixed point solution of the Higgs Mixing model including singlets}
\vspace{\baselineskip}

For the Higgs Mixing model corresponding to the superpotential in
Eq.\ref{rossW} with
symmetric inputs for the $l,h,k$ matrices at a high scale but including the
singlet couplings, we now present the
solutions to the fixed point equations:
\bea
R^{h_{22}}&=&{\frac {583\,}{216}}-R^{h_{21}}+{\frac
{25\,\,b}{112}}-R^{h_{32}}
\nn \\ 
R^{l_{11}}&=&-{\frac
{25\,\,b}{84}}-{\frac{5\,}{2}}+2R^{l_{32}}+R^{l_{33}}+R^{l_{22}}\nn \\
R^{s_{H_2^{(-2)}}}&=&{\frac
{3\,\,b}{28}}+{\frac{1}{2}}-R^{s_{H_2^{(-1)}}}-R^{r_{H_2^{(-1)}}}\nn \\  
R^{r_{H_2^{(-2)}}}&=&R^{s_{H_2^{(-1)}}}+
R^{r_{H_2^{(-1)}}}\nn \\ 
R^{s_{H_1^{(0)}}}&=&R^{\bar{s}_{H_1^{(1)}}} \nn \\
R^{\bar{s}_{H_1^{(0)}}}&=&{\frac {3\,\,b}{28}}+{\frac
{1}{2}}-R^{\bar{s}_{H_1^{(1)}}}-R^{r_{H_1^{(0)}}} \nn \\ 
R^{s_{H_1^{(-1)}}}&=&
{\frac {3\,\,b}{28}}+{\frac {1}{2}}-
R^{\bar{s}_{H_1^{(1)}}}-R^{r_{H_1^{(0)}}} \nn \\ 
\bar{s}_{H_1^{(-1)}}&=&-R^{r_{H_1^{(-1)}}}
+R^{\bar{s}_{H_1^{(1)}}}+R^{r_{H_1^{(0)}}} \nn \\ 
R^{r_{H_1^{(-2)}}}&=&{\frac {3\,\,b}{28}}+{\frac {1}{2}}+R^{r_{H_1^{(-1)}}}-
R^{\bar{s}_{H_1^{(1)}}}-R^{r_{H_1^{(0)}}} \nn \\ 
R^{s_{H_1^{(-2)}}}&=&-R^{r_{H_1^{(-1)}}}+
 R^{\bar{s}_{H_1^{(1)}}}+R^{r_{H_1^{(0)}}} \nn \\ 
R^{s_{H_2^{(0)}}}&=&R^{\bar{s}_{H_2^{(1)}}} \nn \\ 
R^{s_{H_1^{(1)}}}&=&
{\frac {3\,\,b}{28}}+{\frac {1}{2}}-R^{r_{H_1^{(1)}}}-R^{\bar{s}_{H_1^{(1)}}} \nn \\ 
R^{\bar{s}_{H_1^{(2)}}}&=&{\frac {3\,\,b}{28}}+{\frac {1}{2
}}-R^{r_{H_1^{(1)}}}-R^{\bar{s}_{H_1^{(1)}}} \nn \\ 
R^{r_{H_2^{(1)}}}&=&
R^{\bar{s}_{H_2^{(3)}}}+R^{r_{H_2^{(2)}}}-R^{\bar{s}_{H_2^{(1)}}} \nn \\ 
R^{s_{H_2^{(1)}}}&=&{\frac {3\,\,b}{28}}+{\frac {1}{2}}-
R^{\bar{s}_{H_2^{(3)}}}-R^{r_{H_2^{(2)}}} \nn \\ 
R^{\bar{s}_{H_2^{(2)}}}&=&{\frac {3\,\,b}{28}}+{\frac
 {1}{2}}-R^{\bar{s}_{H_2^{(3)}}}-
 R^{r_{H_2^{(2)}}} \nn \\ 
R^{r_{H_2^{(0)}}}&=&{\frac {3\,\,b}{28}}+{\frac
{1}{2}}-R^{s_{H_2^{(-1)}}}-R^{\bar{s}_{H_2^{(1)}}} \nn \\  
\bar{s}_{H_2^{(-1)}}&=&{\frac {3\,\,b}{28}}+{\frac
{1}{2}}-R^{s_{H_2^{(-1)}}}-R^{r_{H_2^{(-1)}}} \nn \\ 
R^{\bar{s}_{H_2^{(0)}}}&=&
R^{s_{H_2^{(-1)}}} \nn \\ 
R^{\bar{s}_{H_1^{(3)}}}&=&
R^{s_{H_1^{(2)}}} \nn \\ 
R^{r_{H_1^{(2)}}}&=&R^{r_{H_1^{(1)}}}-R^{s_{H_1^{(2)}}}+
R^{\bar{s}_{H_1^{(1)}}} \nn \\ 
R^{s_{H_2^{(2)}}}&=&R^{\bar{s}_{H_2^{(3)}}} \nn \\ 
R^{h_{11}}&=&{\frac {25\,\,b}{112}}+{\frac {583\,}{216}}-R^{h_{31}}-R^{h_{21}}
\nn \\  
R^{h_{31}}&=&R^{h_{31}} \nn \\ 
R^{k_{31}}&=&{\frac {463\,}{216}}+{\frac {25\,\,b}{112}}-R^{k_{11}}-R^{k_{21}}
\nn \\  
R^{l_{31}}
&=&{\frac {5\,}{2}}-R^{l_{32}}-R^{l_{33}}+{\frac {25\,\,b}{84}} \nn \\ 
R^{l_{21}}&=&{\frac {25\,\,b}{84}}+{\frac {5\,}{2}}-R^{l_{32}}-R^{l_{22}} \nn
\\  
s_{H_2^{(7)}}&=&R^{\bar{s}_{H_2^{(8)}}} \nn \\ 
R^{r_{H_2^{(8)}}}&=&{\frac {3\,\,b}
{28}}+{\frac {1}{2}}-R^{\bar{s}_{H_2^{(8)}}} \nn \\ 
R^{\bar{s}_{H_2^{(6)}}}&=&R^{s_{H_2^{(5)}}} \nn \\ 
R^{r_{H_1^{(8)}}}&=&
{\frac {1}{2}}+{\frac {3\,\,b}{28}}-
R^{\bar{s}_{H_1^{(8)}}} \nn \\ 
R^{s_{H_1^{(7)}}}&=&R^{\bar{s}_{H_1^{(8)}}} \nn \\ 
R^{r_{H_1^{(3)}}}
&=&{\frac {\,b}{28}}+{\frac
{1}{6}}-R^{r_{H_1^{(1)}}}-R^{r_{H_1^{(-1)}}}-R^{r_{H_1^{(5)}}}+R^{\bar{s}_{H_1^{(8)}}}+
R^{\bar{s}_{H_1^{(7)}}} \nn \\ 
R^{\bar{s}_{H_1^{(4)}}}&=&{\frac {\,b}{14}}+{\frac
{1}{3}}+R^{r_{H_1^{(1)}}}-R^{s_{H_1^{(2)}}}+R^{r_{H_1^{(-1)}}}+R^{r_{H_1^{(5)}}}-R^{\bar{s}_{H_1^{(8)}}}-R^{\bar{s}_{H_1^{(7)}}}
\nn 
\\
R^{\bar{s}_{H_1^{(5)}}}&=&{\frac {3\,\,b}{28}}+{\frac{1}{2}}-R^{s_{H_1^{(5)}}}-R^{r_{H_1^{(5)}}} \nn \\  
R^{s_{H_1^{(4)}}}&=&{\frac {3\,\,b}{28}}+{\frac {1}{2}}-R^{s_{H_1^{(5)}}}-
R^{r_{H_1^{(5)}}} \nn \\ 
R^{s_{H_1^{(3)}}}&=&{\frac {\,b}{14}}+{\frac {1}{3}}+R^{r_{H_1^{(1)}}}-R^{s_{H_1^{(2)}}}+R^{r_{H_1^{(-1)}}}+
 R^{r_{H_1^{(5)}}}-R^{\bar{s}_{H_1^{(8)}}}-R^{\bar{s}_{H_1^{(7)}}} \nn \\ 
R^{k_{32}}&=&-R^{k_{33}}+
R^{k_{11}}+R^{k_{21}} \nn \\ 
R^{h_{33}}&=&{\frac {25\,\,b}{112}}+{\frac {583\,}{216
}}-R^{h_{32}}-R^{h_{31}} \nn \\ 
R^{k_{22}}&=&{\frac {25\,
\,b}{112}}+{\frac
{463\,}{216}}-2R^{k_{21}}-R^{k_{11}}+R^{k_{33}} \nn \\ 
R^{s_{H_2^{(6)}}}&=&R^{\bar{s}_{H_2^{(7)}}} \nn \\ 
R^{\bar{s}_{H_2^{(4)}}}&=&{\frac {\,b}{14}}+{\frac {1}{3}}+R^{r_{H_2^{(2)}}}+R^{r_{H_2^{(-1)}}}-R^{\bar{s}_{H_2^{(1)}}}-R^{\bar{s}_{H_2^{(8)}}}+
 R^{r_{H_2^{(5)}}}-R^{\bar{s}_{H_2^{(7)}}} \nn \\ 
R^{s_{H_2^{(3)}}}&=&{\frac {\,b}{14}}+{\frac {1}{3}}+R^{r_{H_2^{(2)}}}+R^{r_{H_2^{(-1)}}}-R^{\bar{s}_{H_2^{(1)}}}-R^{\bar{s}_{H_2^{(8)}}}+R^{r_{H_2^{(5)}}}-
R^{\bar{s}_{H_2^{(7)}}} \nn \\ 
R^{s_{H_1^{(6)}}}
&=&R^{\bar{s}_{H_1^{(7)}}} \nn \\ 
R^{r_{H_1^{(7)}}}&=&{\frac {3\,\,b}{28}
}+{\frac {1}{2}}-R^{\bar{s}_{H_1^{(8)}}}-R^{\bar{s}_{H_1^{(7)}}} \nn \\ 
R^{\bar{s}_{H_1^{(6)}}}&=&R^{s_{H_1^{(5)}}} \nn \\ 
R^{r_{H_1^{(6)}}}&=&{\frac {3\,\,b}{28}}+{\frac {1}{2}}-R^{s_{H_1^{(5)}}}-R^{\bar{s}_{H_1^{(7)}}} \nn \\ 
R^{r_{H_1^{(4)}}}&=&-{\frac {\,b}{14}}-{\frac {1}{3}}+R^{\bar{s}_{H_1^{(8)}}}+
 R^{\bar{s}_{H_1^{(7)}}}-R^{r_{H_1^{(1)}}}+s_{H_1^{(2)}}-R^{r_{H_1^{(-1)}}}+
R^{s_{H_1^{(5)}}} \nn \\ 
R^{r_{H_2^{(3)}}}&=&{\frac {\,b}{28}}+{\frac {1}{6}}-R^{\bar{s}_{H_2^{(3)}}}-R^{r_{H_2^{(2)}}}-
R^{r_{H_2^{(-1)}}}+R^{\bar{s}_{H_2^{(1)}}}+R^{\bar{s}_{H_2^{(8)}}}-R^{r_{H_2^{(5)}}}+
 R^{\bar{s}_{H_2^{(7)}}} \nn \\ 
R^{s_{H_2^{(4)}}}&=&{\frac {3\,\,b}{28}}+{\frac {1}{2}}-R^{s_{H_2^{(5)}}}-R^{r_{H_2^{(5)}}} \nn \\ 
R^{\bar{s}_{H_2^{(5)}}}&=&{\frac {3\,\,b}{28}}+{\frac {1}{2}}-R^{s_{H_2^{(5)}}}-R^{r_{H_2^{(5)}}} \nn \\ 
R^{r_{H_2^{(4)}}}&=&-{\frac {\,b}{14}}-{\frac {1}{3}}-R^{r_{H_2^{(2)}}}-
R^{r_{H_2^{(-1)}}}+R^{\bar{s}_{H_2^{(1)}}}+R^{\bar{s}_{H_2^{(8)}}}+R^{s_{H_2^{(5)}}}+
R^{\bar{s}_{H_2^{(7)}}} \nn \\ 
R^{r_{H_2^{(6)}}}&=&{\frac {3\,\,b}{28}}+
{\frac {1}{2}}-R^{s_{H_2^{(5)}}}-R^{\bar{s}_{H_2^{(7)}}} \nn \\ 
R^{r_{H_2^{(7)}}}&=&{\frac {3\,\,b}{28}}+{\frac {1}{2}}-
R^{\bar{s}_{H_2^{(8)}}}-R^{\bar{s}_{H_2^{(7)}}}. \label{rossin}
\eea

\vspace{\baselineskip}
\noindent{\large \bf Appendix 4: 
Fixed point solution of Quark-line mixing model including singlets}
\vspace{\baselineskip}

On solving the fixed point Eqs.\ref{rh33}-\ref{modelfixedpointconds}
including the singlet sector, we obtain
\bea
R^{s_{Q^{(-1)}}}&=&-{\frac
{95\,b}{1078}}-{\frac {13865}{14553}}- R^{r_{Q^{(-1)}}}- R^{s_{Q^{(-2)}}}\nn
\\  R^{\bar{s}_{Q^{(-1)}}}&=&{\frac 
{115}{441}}+{\frac {3\,b}{98}}+ R^{s_{Q^{(-2)}}}\nn \\  R^{\bar{s}_{Q_{2}}}&=&-{\frac
{117\,b}{539}}-{\frac {10930}{4851}}- R^{r_{Q^{(0)}}}+ R^{r_{Q^{(-1)}}}+
R^{s_{Q^{(-2)}}}\nn
\\ 
R^{\bar{s}_{Q_{3}}}&=&{\frac {575}{1323}}+{\frac {5\,b}{98}}-
R^{\bar{s}_{Q^{(0)}}}- R^{r_{Q^{(-1)}}}- R^{s_{Q^{(-2)}}}\nn \\  
R^{s_{Q_{2}}}&=&{\frac {267\,b}{1078}}+{\frac
{1355}{539}}+  
R^{r_{Q^{(0)}}}- R^{r_{Q^{(-1)}}}- R^{s_{Q^{(-2)}}}\nn \\  R^{s_{Q_{3}}}&=&-{\frac
{5465}{4851}}+R^{h_{(0)3}} 
-{\frac {117\,b}{1078}}+ R^{\bar{s}_{Q^{(0)}}}+ R^{r_{Q^{(-1)}}}+ R^{s_{Q^{(-2)}}}\nn \\ 
R^{s_{Q^{(3)}}}&=&{\frac {5465}{4851}}+{\frac {117\,b}{1078}}+
R^{\bar{s}_{Q^{(4)}}}\nn \\R^{\bar{s}_{Q^{(3)}}}&=&-{\frac {6\,b}{77}}-{\frac {200}{231}}- R^{\bar{s}_{Q^{(4)}}}-  
R^{r_{Q^{(3)}}}\nn \\  R^{r_{Q^{(4)}}}&=&-{\frac {10070}{14553}}-{\frac {31\,b}{539}}-  
R^{\bar{s}_{Q^{(4)}}}\nn \\  R^{s_{Q^{(2)}}}&=&{\frac
{5\,b}{98}}+{\frac {575}{1323}}- R^{\bar{s}_{Q^{(4)}}}- R^{r_{Q^{(3)}}}\nn \\  
R^{h_{(-1)2}}&=&{\frac {13865}{14553}}+{\frac {95\,b 
}{1078}}\nn \\
R^{h_{(4)1}}&=&{\frac {13865}{14553}}+{\frac {95\,b}{1078}}\nn \\ 
R^{s_{Q_{1}}}&=& 
{\frac {3\,b}{98}}+{\frac {115}{441}}\nn \\ 
R^{s_{Q^{(0)}}}&=&{\frac {115}{441}}-R^{h_{(0)3}}- R^{r_{Q^{(0)}}}-
R^{\bar{s}_{Q^{(0)}}}+{\frac 
{3\,b}{98}}\nn \\R^{h_{33}}&=&-R^{h_{(0)3}}+{\frac {13865}{14553}}+{\frac
{95\,b}{1078}}\nn \\ 
R^{\bar{s}_{Q^{(1)}}}&=&{\frac{192\,b}{539}}+{\frac {17660}{4851}}\nn \\
R^{r_{Q^{(2)}}}&=&{\frac {115}{1323}}+{\frac {b}{98}}- R^{r_{Q^{(0)}}}+
R^{\bar{s}_{Q^{(4)}}}- R^{r_{Q^{(-3)}}}+ R^{s_{Q^{(-2)}}}\nn \\ 
R^{s_{Q^{(1)}}}&=&-{\frac {139\,b}{1078}}-{\frac
{18925}{14553}}+ R^{r_{Q^{(-1)}}}+ R^{r_{Q^{(3)}}}+ R^{r_{Q^{(-3)}}}\nn \\
R^{r_{Q^{(1)}}}&=&{\frac{86\,b}{539}}+{\frac {22720}{14553}}-
R^{r_{Q^{(-1)}}}- R^{r_{Q^{(3)}}}- 
R^{r_{Q^{(-3)}}}\nn \\ 
R^{\bar{s}_{Q^{(-3)}}}&=&{\frac {b}{98}}+{\frac {115}{1323}}- R^{r_{Q^{(-3)}}}+ R^{s_{Q^{(-2)}}}+ 
 R^{r_{Q^{(-2)}}}\nn \\  R^{r_{Q^{(-4)}}}&=&{\frac {460}{1323}}+{\frac {2\,b}{49}}+
R^{r_{Q^{(-3)}}}- R^{s_{Q^{(-2)}}}- R^{r_{Q^{(-2)}}}\nn \\
R^{s_{Q^{(-3)}}}&=&{\frac{b}{49}}+{\frac{230}{1323}}- R^{s_{Q^{(-2)}}}-
R^{r_{Q^{(-2)}}}\nn \\
R^{\bar{s}_{Q^{(-2)}}}&=&{\frac {3\,b}{98}}+{\frac {115}{441}}-
R^{s_{Q^{(-2)}}}- R^{r_{Q^{(-2)}}}\nn \\  R^{s_{Q^{(-4)}}}&=&-{\frac{b}{98}} 
-{\frac {115}{1323}}- R^{r_{Q^{(-3)}}}+ R^{s_{Q^{(-2)}}}+ R^{r_{Q^{(-2)}}}\nn
\\  
R^{\bar{s}_{Q^{(2)}}}&=&-{\frac {3\,b}{98}}-{\frac {115}{441}}+ R^{r_{Q^{(0)}}}+ R^{r_{Q^{(3)}}}+ R^{r_{Q^{(-3)}}}- R^{s_{Q^{(-2)}}}.\label{solns}
\eea 
Eq.\ref{solns} shows that in fact 25 out of the 34 variables are
constrained. This must mean that within
Eqs.\ref{rh33}-\ref{modelfixedpointconds}, 9 of the 34 constraints exhibit
degeneracy. Note that for $R^{r_{Q^{(4)}}} > 0$, we must pick
$R^{\bar{s}_{Q^{(4)}}}$ to be negative.

\end{document}